# Recent Advances in Programmable Photonic-Assisted Ultrabroadband Radio-Frequency Arbitrary Waveform Generation

A. Rashidinejad, *Member, IEEE*, Y. Li, *Member, IEEE*, and A. M. Weiner, *Fellow, IEEE*

*Abstract*— This paper reviews recent advances in photonic-assisted radio-frequency arbitrary waveform generation (RF-AWG), with emphasis on programmable ultrabroadband microwave and millimeter-wave waveforms. The key enabling components in these techniques are programmable optical pulse shaping, frequency-to-time mapping via dispersive propagation, and high-speed photodetection. The main advantages and challenges of several different photonic RF-AWG schemes are discussed. We further review some proof-of-concept demonstrations of ultrabroadband RF-AWG applications, including high-resolution ranging and ultrabroadband non-line-of-sight channel compensation. Finally, we present recent progress toward RF-AWG with increased time aperture and time-bandwidth product.

*Index Terms*— Radio-frequency photonics, Microwave and millimeter-wave generation, Arbitrary waveform generation, Optical pulse shaping, Frequency-to-time mapping, Modulation.

## I. INTRODUCTION

ULTRABROADBAND radio-frequency (RF) waveforms are of interest for numerous applications, such as high-speed communication, biomedical imaging and tomography, chemical spectroscopy and high-resolution ranging [1]–[13]. As a result of the extreme congestion of low-frequency bands of the RF spectrum, the evolution of these applications is contingent upon successful waveform generation and processing techniques within the high-microwave and millimeter-wave (MMW) frequency regions [14]–[18]. In this regards, state-of-the-art electronics suffers from limited digital-to-analog conversion speeds, which in turn constrains the frequency range and bandwidth of such systems. For example, the operating frequency range of currently available RF-arbitrary waveform generation (RF-AWG) systems based on electronics is confined to DC–20 GHz [19]. Upconversion [5], [20], [21] of electronically generated arbitrary baseband signals to higher frequency bands also suffers bandwidth constraints, bounded by the intermediate frequency (IF) bandwidth of the mixer. Ultrabroadband electronic systems are further challenged by issues like high timing jitter, susceptibility to electromagnetic interference, high cabling loss, and large size [16]–[18]. Substantial research into photonic-assisted generation of high-frequency and ultrabroadband RF arbitrary waveforms (RF-AWG) has been conducted to address these shortcomings [22]–[26].

In photonic-assisted RF-AWG techniques, a short optical input pulse is first shaped in the optical domain and then converted into microwave or MMW frequencies through high-speed photodetection [26]–[28]. These schemes can theoretically generate RF waveforms with much larger center frequencies and bandwidths compared to their electronic counterparts, while also being compatible with radio-over-fiber technology for convenient and low-loss signal distribution. Additionally, recent experiments have demonstrated initial steps toward on-chip integration of such systems [29]–[35]. Many methods for photonic-assisted RF-AWG have been discussed, including direct space-to-time (DST) pulse shaping [36]–[41]; spectral-domain pulse shaping [42]–[44] and frequency-to-time mapping (FTM) [45]–[60]; as well as other schemes [61]–[68].

Among the many photonic RF-AWG techniques, those relying on Fourier transform (FT) optical pulse shaping followed by FTM offer advantages in terms of programmability and attainable signal complexity [45], [46], [59]. In these techniques the spectrum of an ultrashort optical input pulse is first tailored and then mapped to the time domain by stretching the optical pulses through a dispersive medium, resulting in an optical intensity profile that is scaled version of the tailored spectrum. The optical intensity waveform is then mapped to the RF domain using a high-speed photodetector. Although spectral tailoring techniques based on fiber grating structures or on-chip optical pulse shapers have been demonstrated for RF-AWG [30]–[33], [37], in most implementations spectral shaping is carried out using bulk optic FT pulse shapers [42]–[44]. These subsystems provide programmable, independent manipulation of the spectral amplitude and phase of an input

Manuscript received September 21, 2015. The authors are with the School of Electrical and Computer Engineering, Purdue University, West Lafayette, IN 47907-2035 USA. E-mail: {arashidi, li592, amw} @purdue.edu. This work was supported in part by the Office of the Assistant Secretary of Defense for Research and Engineering under the National Security Science and Engineering Faculty Fellowship program through grant N00244-09-1-0068 from the Naval Postgraduate School.

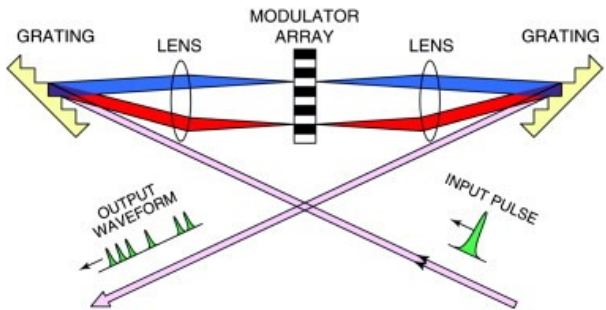

Fig. 1. Basic setup for Fourier transform optical pulse shaping [44].

optical waveform using spatial light modulators, usually based on liquid crystal technology. In addition to traditional home-built setups, commercial FT pulse shapers (also known as waveshapers) based on liquid crystal on silicon (LCoS) technology have become popular in recent years [69], [70].

In this article we first review the basics of several different photonic-assisted RF-AWG techniques as well as some of our group's most recent advances (section II). Special emphasis is placed on programmable RF-AWG, which generally utilizes programmable optical pulse shapers. A great deal of work exploring other schemes, such as shaping via novel fiber Bragg grating (FBG) structures, has also been reported; however, in most cases such work results in waveforms that are either fixed or only slightly tunable. Such work will be covered only briefly in this review. Programmable photonic-assisted RF-AWG has matured to the point that it can be exploited in proof-of-concept application experiments. One application that received considerable attention several years ago in our group concerns compensation of the strong frequency-dependent delay exhibited by certain broadband antennas even in simple line-of-sight configurations. Experiments demonstrating such compensation are discussed in a review of photonic-assisted RF-AWG published as a book chapter in 2013 [26]. Here we summarize more recent examples, including ranging experiments in the W band (75-110 GHz) that achieve few millimeter range resolution and experiments demonstrating characterization and compensation of ultrawideband (UWB), non-line-of-sight, indoor wireless communication channels dominated by strong multipath distortion (section III). One issue affecting most work in photonic-assisted RF-AWG is that waveforms may be generated over only a limited time aperture. In section IV we present our recent progress in circumventing such time aperture limits via photonics-based pseudorandom phase-coding of a generated RF waveform sequence that would otherwise repeat periodically. This enables an increase in the nanosecond-level time aperture achieved by conventional photonic-assisted RF-AWG schemes to several microseconds and beyond, with corresponding increase in time-bandwidth product.

## II. PHOTONICS-ASSISTED RF-AWG

The availability of well-developed techniques for shaping ultrashort light pulses and generating optical arbitrary waveforms [27], [28], [42]–[44] forms the basis for photonic-assisted RF-arbitrary waveform generation (RF-AWG).

Starting roughly fifteen years ago, a wide variety of RF-AWG schemes have been presented and experimentally explored. One early experiment from our group utilized FT pulse shaping and a photoconductive dipole antenna to generate programmable terahertz bursts (ca. 1 THz center frequency) but was limited to very simple waveform shapes [71]. An interesting point revealed by this early experiment was the possibility of enhancing the peak power spectral density of the emitted THz by shaping the optical excitation to avoid photodetector saturation by reducing peak intensity. A similar concept was later employed to enhance low timing jitter signal generation in the low GHz range via photodetection of stabilized optical frequency combs [72]. The basic setup of an FT optical pulse shaper, utilized in [71] and also extensively adopted in later photonic RF-AWG techniques, is displayed in Fig. 1. The key idea is that the incident optical pulse is first decomposed into its constituent spectral components by a spectral disperser (usually a grating) and a focusing element (a lens or a curved mirror). A spatially patterned mask (usually a programmable spatial light modulator (SLM)) then modulates the phase and amplitude, and sometimes the polarization, of the spatially dispersed spectral components. After the spectral components are recombined by a second lens and grating, a shaped output pulse is obtained, with the pulse shape given by the Fourier transform of the pattern transferred by the mask onto the spectrum [44].

In this section, we review results from several different RF-AWG schemes. Our discussion is organized as follows: (II.A) direct space-to-time (DST) pulse shaping; (II.B) Fourier transform pulse shaping and frequency-to-time mapping (FTM); (II.C) advanced FTM-based techniques; (II.D) steps toward integrated solutions for RF-AWG.

### A. Direct Space-to-Time Mapping

One of the earliest optical pulse shaping techniques that was employed for RF-AWG is direct space-to-time (DST) pulse shaping [36]–[41]. Unlike FT shaping, in DST pulse shaping the spatial mask or SLM is placed adjacent to the spectral disperser (diffraction grating). The output waveform, which is collected through a narrow slit (or optical fiber) placed at the back focal plane of a single lens following the grating, corresponds to the input spatial profile directly scaled into the time domain [36]. By appropriate tailoring of the spatial mask, the output optical waveform can be correspondingly adjusted. To convert to the RF domain, one simply employs a photodetector (PD), as in other RF-AWG schemes.

In [38] and [39], by using a DST pulse shaping configuration enhanced to produce waveforms over larger than usual time aperture, microwave and MMW arbitrary waveform generation up to 50 GHz was for the first time accomplished. By appropriately tailoring the optical pulse sequences driving the photodetector, the ability to synthesize strongly phase- and frequency-modulated MMW electrical signals on a cycle-by-cycle basis was demonstrated. Fig. 2(a) shows the results for phase-modulation of a 48 GHz burst waveform. The top trace is an unmodulated, ~48 GHz burst generated using DST pulse shaping. In the lower traces, a $\pi$ phase shift is injected after the second, third, and fourth cycle in the electrical waveform, respectively, as determined by the location of an extra 10-ps delay in the driving optical pulse sequence. The top curve of Fig. 2(b) shows an example of frequency modulation. The first cycles of the waveform occur at 48 GHz, which then changes abruptly to 24 GHz for the last two cycles. Here the frequency modulation is controlled by variation of the repetition rate of the driving optical pulse sequence in concert with the RF filtering that occurs as a result of electrical bandwidth limitations. The dashed curve in this figure is an ideal 48/24 GHz waveform, which differs only slightly from the generated waveform. A cross-correlation measurement of the optical waveform sequence driving the PD is also shown (lower curve of Fig. 2(b)). This example illustrates that even crude shaping in the optical domain can result in realization of high quality waveforms in the RF domain after the smoothing that results from the limited electrical bandwidth (here ~50 GHz). The waveform examples shown in Fig. 2 remain beyond the capabilities of the highest bandwidth electronic arbitrary waveform generators available today.

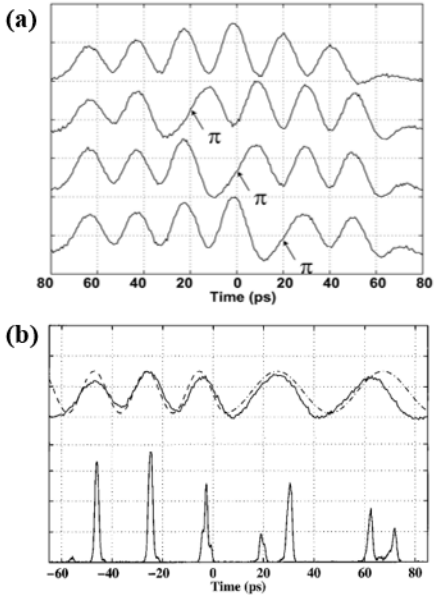

Fig. 2. Initial experimental results for direct space-to-time mapping RF-AWG – (a) Phase- and (b) frequency-modulated MMW experimental results using a DST-based RF-AWG setup [38], [39]. The lower curve of Fig. 2(b) is the cross-correlation measurement of the optical sequence driving the photodetector.

The main disadvantages of DST-based optical shaping are bulkiness, high loss, and low degree of complexity of the generated waveforms. DST-based optical pulse shaping was realized in a modified scheme using integrated optic arrayed-waveguide grating spectral dispersers [37]. A recent RF-AWG scheme based on a related time domain shaping approach implemented in a silicon photonics chip [33] is briefly discussed in section II-D.

*B. Fourier Transform Pulse Shaping and Frequency-to-Time Mapping (FTM)*

As previously mentioned, photonic-assisted RF-AWG techniques based on Fourier transform (FT) optical pulse shaping and FTM are particularly powerful and offer better performance in terms of generated signal complexity. Fig. 3(a) shows the generic layout of a spectral pulse shaping and FTM-based RF-AWG scheme, introduced by [45], [46]. These techniques implement arbitrary signal tailoring in the optical frequency domain, often using a programmable FT pulse shaper [42]–[44]. This class of photonic RF-AWG schemes has received sustained attention from the research community. In these schemes, appropriately shaped optical pulses are intentionally subjected to strong dispersive propagation, which results in an optical intensity profile that is a scaled replica of the shaped optical spectrum. The frequency-to-time mapping phenomenon is nicely depicted in Fig. 3(b). As represented in this plot, when the shaped spectrum propagates through dispersion, different wavelengths travel at different speeds (only 4 wavelengths are shown for simple illustration). For sufficiently large chromatic dispersion, referred to as far field condition [55], a linear mapping of the input power spectrum to the output temporal intensity profile is achieved. The resultant optical intensity is then mapped to the electrical domain using a high-speed PD.

Thus in FTM-based RF-AWG satisfying the far field condition, the generated RF waveform $v_{RF}(t)$ is expressed as:

$$v_{RF}(t) \propto |a_{out}(t)|^2 = \left|A_{in}\left(\omega = -\frac{t}{\psi_2}\right)\right|^2 \quad (1)$$

where, according to Fig. 3, $\psi_2$ represents the dispersion (units $ps^2$); $a_{in}(t)$ and $a_{out}(t)$ are the complex temporal envelope functions of the optical field before and after dispersive propagation; and $A_{in}(\omega)$ is the Fourier transform of the shaped waveform $a_{in}(t)$.

We observe that the complexity of the programmable waveforms that can be generated using these RF-AWG schemes depends directly on the complexity of the spectral filter function that can be imprinted on the optical pulses via pulse shaping. Pulse shaping can be implemented using SLM-based optical pulse shapers [45]–[48], [55]–[60], fiber Bragg grating (FBG) structures [49]–[54], arrayed-waveguide grating processors [30], [31], [73], and ring resonator-based programmable optical filters [31]–[34]. Fig. 4 shows the setup and experimental results for one of the early demonstrations of FTM-based RF-AWG for ultra-wideband (UWB) waveform generation [46], [47]. UWB technology (3.1 – 10.6 GHz) is a baseband communication scheme that has been extensively studied due to its potential for short range, high data rate wireless communications and ranging [74], [75]. Since UWB signals have noise-like RF spectra, with power spectral density regulated by the FCC at or below -41.3 dBm/MHz, UWB systems may be overlaid with minimal interference with existing, regulated narrowband wireless services. In the experiments of Fig. 4, transform-limited ultrashort optical pulses from a mode-locked (ML) laser are shaped in a reflective Fourier transform optical pulse shaper, built using a 128 pixel liquid crystal modulator (LCM) array, and subsequently dispersed in ~5.5 km of single-mode fiber (SMF) and detected with a 22 GHz photodiode. Figs. 4(b) and 4(c) depict an example of an abruptly frequency-hopped UWB signal; the similarity between the shaped optical spectrum and the generated RF time domain waveform is clearly evident. In [46] our group investigated the use of this RF-AWG approach for spectral engineering within the UWB band. It was shown that the spectral content of the generated waveforms could be precisely "engineered" to conform to FCC power spectral density regulations. Figs. 4(d)-4(f) show waveforms with detailed structure designed in order to generate, for the first time, programmable RF waveforms with nearly constant power spectral density spanning the full UWB frequency range.

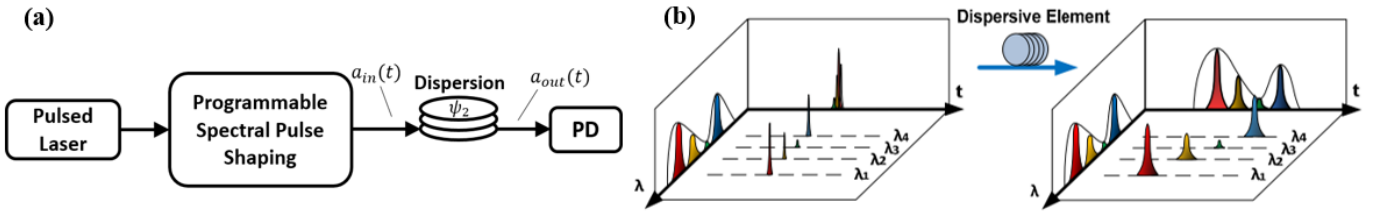

Fig. 3. RF-AWG based on optical spectral pulse shaping and frequency-to-time mapping – (a) Schematic diagram of the generic setup. (b) Illustration of frequency-to-time mapping in a dispersive element (adapted from [56]).

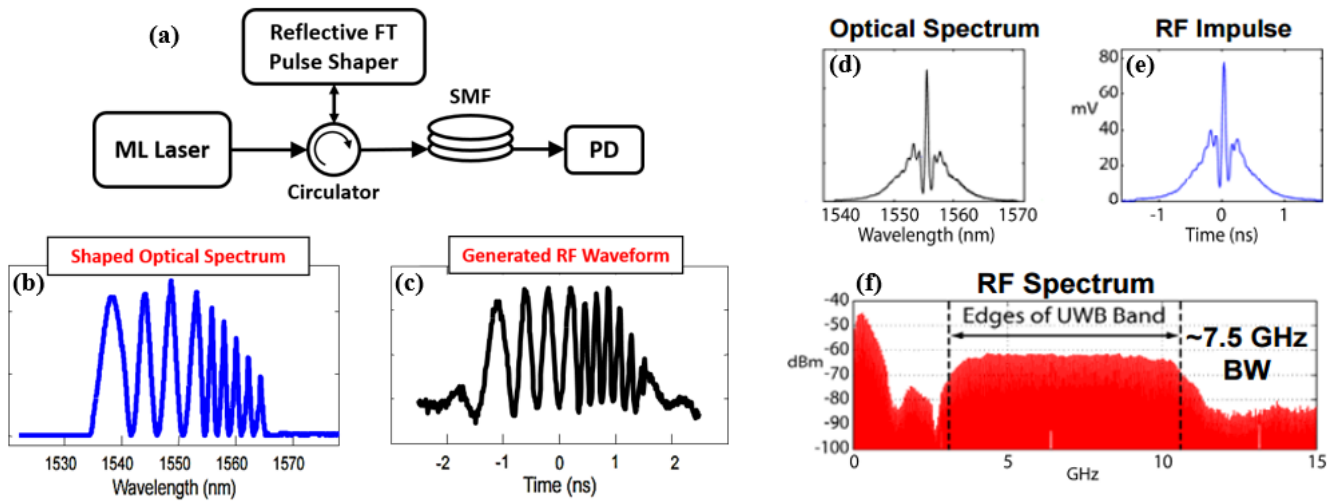

Fig. 4. FTM-based microwave arbitrary waveform generator using a reflective optical Fourier transform pulse shaper – (a) Setup schematic. (b),(c) Experimental demonstration of FTM for a UWB linear frequency downchirp [46]. (d)-(f) Experimental results for spectral engineering spanning the full UWB frequency range (3.1 GHz to 10.6 GHz) [47].

As observed in Fig. 4, the generated RF waveforms have an inevitable DC component, due to the fact that conventional photodetectors respond to the (non-negative) optical intensity. However, by slightly altering the generation setup and incorporating balanced photodetection, McKinney was able to demonstrate background-free RF-AWG [76]. In this scheme programmable optical intensity waveforms exhibiting complementary amplitude apodization are synthesized via polarization pulse shaping. This is done by orienting the axis of the LCM in the utilized pulse shaper at 45° with respect to the incoming optical polarization. Balanced photodetection is then utilized after dispersive propagation to suppress the waveform pedestal in the electrical domain. Fig. 5 depicts the time-domain and spectral-domain results for UWB chirped waveform generation within 4-6 GHz using this scheme. This work clearly demonstrates successful background removal, resulting in more than 25 dB suppression of unwanted low-frequency content. Balanced detection also improves the generated RF power by 6 dB as seen in Fig. 5(b).

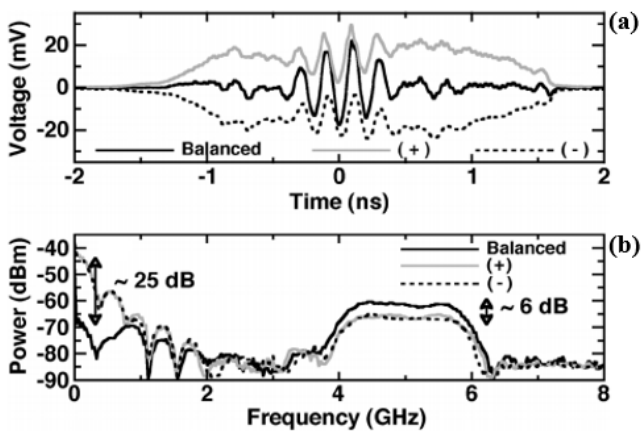

Fig. 5. Background-free RF-arbitrary waveform generation [76] – Experimental results for broadband UWB chirp generation, obtained via single-ended (positive: gray, negative: dashed black) and balanced (solid black) photodetection in (a) time-domain and (b) frequency domain. MLFL: mode-locked fiber laser, PMF: polarization-maintaining fiber.

A different approach in FTM-based RF-AWG techniques is to utilize specially-engineered fiber Bragg gratings (FBG) for the pulse shaping and/or FTM stages [49]–[54]. Spatially-discrete chirped FBGs (SD-CFBG) can be fabricated to provide an optical filter response for spectral shaping, replacing the FT pulse shaper; and/or frequency-dependent delay to replace the relatively long fiber in the FTM stage. An example that utilizes SD-CFBGs for both roles simultaneously is reported in [53] and plotted in Fig. 6(a). In this design, the SD-CFBG is fabricated to impart the relevant amplitude filter function for microwave generation, while having a large dispersive response for FTM. Figs. 6(b) and 6(c) depict experimental results from this setup for generating linear and nonlinear downchirped MMW waveforms with frequency content up to ~40 GHz and time-bandwidth products of up to 23.2. Time-bandwidth product (TBWP), which is the product of the generated RF waveform's time aperture and its positive bandwidth (typically at 3-dB), is a very important criterion of an RF-AWG technique and determines the versatility of the scheme in creating a wide variety of complex waveforms. A large TBWP is also important from the application perspective. For example, in MMW imaging, the key figure-of-merit in evaluating a spread spectrum MMW waveform is its compression ratio, which determines the achievable imaging resolution. This compression ratio is directly proportional to the waveform's time-bandwidth product.

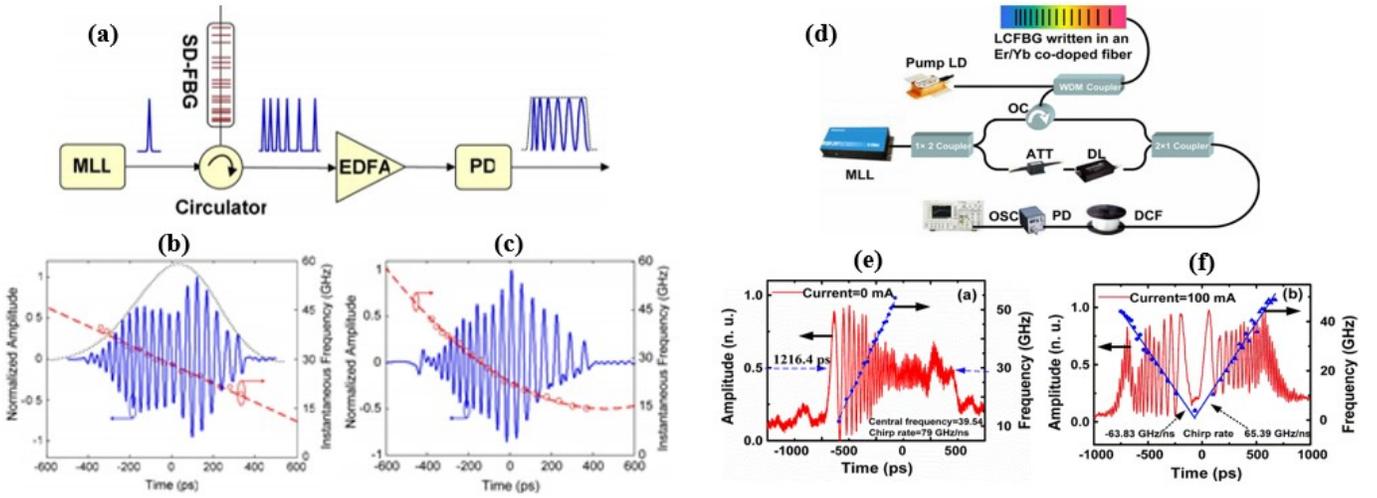

Fig. 6. Fiber Bragg grating-based RF waveform generation – (a) MMW chirp generation setup using SD-CFBG [53], and the generated (b) linear and (c) nonlinear chirps along with instantaneous frequency. (d) MMW tunable linear chirp generation using optically-pumped LCFBG [54], and (e), (f) the generated chirps with different center frequencies and chirp rates. MLL: mode-locked laser, DCF: dispersion compensating fiber.

Although CFBG-based schemes are very compact and allow RF waveform generation with relatively large TBWP, they suffer from lack of programmability. That is, once the CFBG is engineered, the generated waveform cannot be altered and thus, in a sense, RF arbitrariness is sacrificed. However, recent research in this area has shown that by incorporating optically-pumped CFBG structures, a certain amount of RF waveform programmability is attainable [28], [29]. An example of MMW chirp generation with tunable chirp rate and center frequency is reported in [54]. The experimental setup for this scheme is plotted in Fig. 6(d), where pulses from a mode-locked laser are directed to an optical interferometer with an optically-pumped linear CFBG (LCFBG) in one arm and tunable optical delay line (DL) and attenuator (ATT) in the other. The LCFBG imparts a quadratic phase onto the optical spectrum that can be slightly changed by altering the pump power from the laser diode (LD). From the RF waveform perspective, this results in MMW linear chirp generation with tunable chirp rate. Furthermore, through a process discussed further in section II-C, varying the optical delay allows tuning of the center frequency of the MMW chirp. Figs. 6(e) and 6(f) show two examples of results obtained from this setup. To obtain the linear upchirp of Fig. 6(e), the current to the LD is set to zero, while the DL is tuned to achieve 39.5 GHz center frequency; whereas for Fig. 6(f), the LD is driven at 100 mA and the DL is tuned so that the central portion of the waveform corresponds to zero frequency. As a result of the different pump powers to the CFBG, the chirp rate is changed from ~79 GHz/ns to ~65.39 GHz/ns.

### C. Advanced FTM-Based Techniques

Previously mentioned FTM-based RF-AWG schemes assume operation in the far field, i.e., the magnitude of the dispersion is large enough that Equation (1) is satisfied. The far field condition is closely analyzed in [55] by Torres-Company et al. Propagating an optical field ($a_{in}(t)$) through a quadratic dispersive element ($\psi_2$) results in:

$$a_{out}(t) \propto e^{-j\left(\frac{t^2}{2\psi_2}\right)} \int_{\langle \sigma_{in} \rangle} a_{in}(t') e^{-j\left(\frac{t'^2}{2\psi_2}\right)} e^{j\left(\frac{tt'}{\psi_2}\right)} dt' \qquad (2)$$

where $\langle \sigma_{in} \rangle$ is the time aperture of the input optical pulse (units *ps*). For Equation (1) to be satisfied, the quadratic phase term inside the integral in Equation (4) must be negligible over the duration of $a_{in}(t)$, i.e. in the simplest form we must have:

$$|\psi_2| > \sigma_{in}^2/\pi \qquad (3)$$

Ref. [55] shows that working in the far-field region limits the maximum RF frequency that can be generated using the RF-AWG setup. As a result, photonic-assisted RF-AWG using FTM-based schemes could not extend to higher frequency regime (e.g. 40 GHz) without sacrificing waveform fidelity. In 2013, Dezfooliyan et al introduced a new scheme called near-field FTM (NF-FTM) that overcame the restrictive far field requirement [56]. Unlike conventional FTM, when the far-field condition does not hold, the output optical intensity depends on both the amplitude and the phase of the input optical spectrum. In NF-FTM, the input spectrum is predistorted in both amplitude and phase so that the desired output intensity in time is achieved after dispersive propagation. If $a_{FTM}(t)$ refers to the inverse Fourier transform of $A_{in}(\omega)$, according to (1), that is required to be programmed onto the spectrum assuming the far-field condition is satisfied, in NF-FTM, the complex spectrum corresponding to $a_{NF-FTM}(t)$ is programmed, which is defined as follows:

$$a_{NF-FTM}(t) = a_{FTM}(t) \cdot \exp(jt^2/2\psi_2) \quad (4)$$

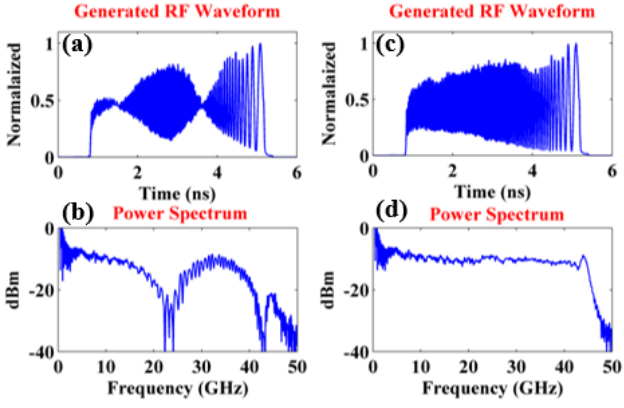

Fig. 7. Generating ultrabroadband down-chirp RF waveform from DC to ~45 GHz with time aperture of ~4.5 ns, corresponding to a TBWP of ~200 – (a-b) Waveforms from conventional FTM. Generated RF waveform is badly distorted and certain frequencies are strongly attenuated. (c-d) Waveforms from near-field FTM. A beautiful chirp is obtained and the RF spectrum extends smoothly out to ~45 GHz [56].

By substituting (4) in (2) it is clear that the resultant output optical waveform would resemble the desired waveform, regardless of the length of dispersion. Fig. 7 shows experimental results for generation of an ultrabroadband down-chirp signal with frequency content from nearly DC to 45 GHz. For this waveform, the far field requirement is strongly violated. For the conventional FTM scheme, the generated RF waveform is badly distorted (Fig. 7(a)), and certain groups of frequencies are strongly attenuated (Fig. 7(b)). For NF-FTM, however, after dispersive propagation a well-behaved ultrabroadband RF waveform is obtained (Fig. 7(c)), in close agreement with the ideal waveform (not shown here). The RF spectrum (Fig. 7(d)) extends smoothly out from DC to ~45 GHz, more than a factor of two beyond the highest frequency available from commercial electronic arbitrary waveform generators. This combination of high RF bandwidth and large TBWP, while maintaining excellent waveform shape and fidelity, had not been previously obtained in photonic-assisted or electronic RF-AWG.

Despite the impressive results demonstrated with NF-FTM, TBWP is reduced when NF-FTM is applied for generation of passband waveforms, e.g., waveforms fitting within the unlicensed 57-64 GHz band for short range wireless or in the W-band (75-110 GHz) utilized in radar applications [56], [59]. This is because in NF-FTM, as well as conventional FTM-based RF-AWG, the complete temporal profile of the desired waveform, including high frequency fringes, must be programmed onto the pulse shaper, which is why these schemes are also referred to as baseband RF-AWG. In 2014, we presented an advanced FTM-based RF-AWG scheme that overcame these limitations to provide programmable microwave, MMW and even sub-THz generation with broadly tunable center frequency while preserving maximal TBWP capability even for strongly banded signals [59]. The idea was to utilize a delay-mismatched optical interferometer to form the oscillations corresponding to the center frequency of the RF passband, thereby allowing the pulse shaper to devote all of its resources (i.e. programmable pixels for phase and amplitude control) to tailor the complex envelope of the desired waveform. The setup for this technique is depicted in Fig. 8(a): mode-locked laser pulses are directed into a delay-mismatched interferometer with a programmable pulse shaper in one arm, followed by dispersive propagation for FTM. The passband RF arbitrary waveform generated in this setup can be characterized as follows:

$$v_{RF}(t) \propto \left|H\left(\frac{-t}{\psi_2}\right)\right| \cdot \cos\left(\frac{t\tau}{\psi_2} - \angle H\left(\frac{-t}{\psi_2}\right) + \phi_0\right) \quad (5)$$

where $H(\omega) = |H(\omega)| \angle H(\omega)$ is the complex pulse shaper transfer function, $\tau$ is the delay imbalance, and $\phi_0$ is a deterministic phase term [59]. This scheme provides independent control over the temporal amplitude, temporal phase, and center frequency of the generated passband waveforms.

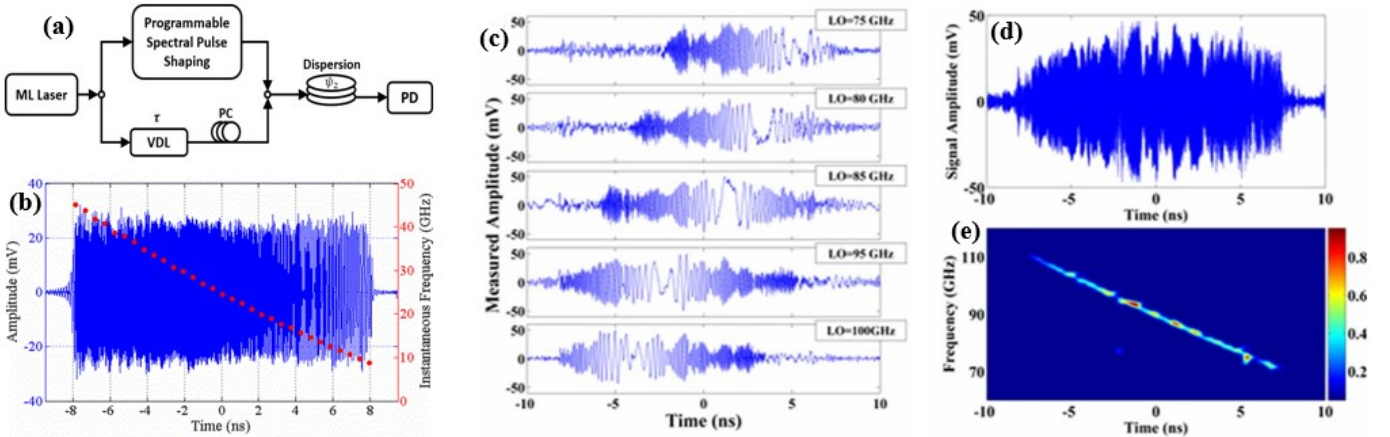

Fig. 8. Interferometric (passband) RF-AWG achieving maximal TBWP for passband microwave, MMW and even sub-THz frequencies – (a) Setup schematic. (b) MMW down-chirp generation spanning 7-45 GHz with TBWP of ~589 and instantaneous frequency plot [59]. (c) W-band (70-110 GHz) ultrabroadband chirp downconversion measurements with various local oscillator (LO) frequencies (limited mixer bandwidth, requires multiple LO measurements), (d) reconstructed W-band waveform and (e) corresponding spectrogram plot. The TBWP of RF-AWG in W-band meets the upper bound of ~600 [60].

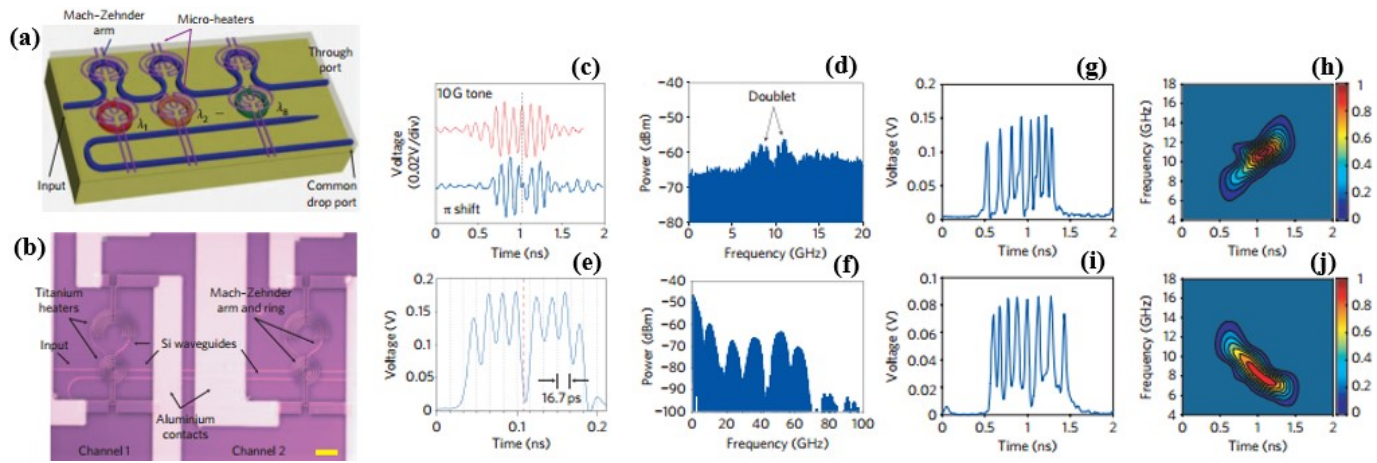

Fig. 9. On-chip spectral pulse shaper implementation using cascaded tunable microring resonators in an MZ configuration for RF-AWG – (a) Cartoon representation and (b) optical image (including 2 rings) of the device. Scale bar, 20mm. (c),(d) Implementing a $\pi$ phase shift in a generated 10 GHz burst. (e),(f) Implementing a $\pi$ phase shift in a generated 60 GHz burst. (g),(h) Broadband RF waveform and spectrogram of an up-chirped signal. (i),(j) Broadband RF waveform and spectrogram of a down-chirped signal [32].

Through mathematical analysis and experimental demonstrations, we validated that the TBWP of this setup is equal to the upper bound set by the number of independent pulse shaper control elements, providing TBWP twice as large as conventional frequency-to-time mapping techniques (including NF-FTM). Fig. 8(b) shows a generated down-chirp waveforms with 16 ns time aperture and frequency content from 7 to 45 GHz, for a TBWP of ~589. Plotted in Figs. 8(c)-8(e) are measurements of a generated down-chirp waveform with ~15 ns time aperture and frequency content from 70 to 110 GHz, for a TBWP of ~600 [60]. To our knowledge the latter experiments are the first to demonstrate RF-AWG covering the full W-band, a frequency region of interest for various broadband applications, such as ultrahigh-speed wireless communications, high-resolution ranging, electromagnetic imaging, and high-speed tomography. In section III-B, we will discuss high-resolution ranging experiments performed using such high TBWP W-band waveforms [60]. The passband nature of the setup sketched in Fig. 8(a) should enable RF-AWG at even higher frequencies, up to several hundreds of gigahertz, limited only by the speed of the photodetector.

### D. Integrated Solutions

Scaling down the size of the setups through integration would promote potential applications of photonic-assisted RF-AWG. Several steps have been made in this direction using on-chip optical pulse shapers [29]–[35], [73], including FT pulse shapers based on integrated array waveguide grating structures as the spectral dispersers [29]–[31], [73]. In this subsection, we describe two works from our group, in which pulse shaping structures using arrays of microring resonators in silicon photonic chips are utilized for the RF-AWG [32], [33].

Fig. 9(a) shows a silicon photonic spectral pulse shaper, implemented using 8 cascaded microring resonators, all in an add–drop configuration [32]. An optical image of 2 rings from the fabricated device is also included in Fig. 9(b). Each microring selectively transfers the optical power at its resonance wavelength from the input waveguide (through-port) to the output waveguide (drop-port). A micro-heater is placed above each ring, which can locally and independently control its resonance frequency. Each ring is also provisioned with a micro-heater tuned interferometric coupling structure, which allows thermal tuning of the maximum transmission to the drop port. By thermal tuning of the resonant frequencies and the coupled power, spectral amplitude shaping can be realized at the output waveguide. The output was sent to ~5.5 km of SMF for FTM, scaling the shaped spectra into nanosecond scale temporal intensity waveforms that are converted into the RF domain via photodetection. Figs. 9(c)-9(f) show temporal and spectral domain plots of waveform bursts centered at 10 GHz and 60 GHz, respectively, each incorporating an abrupt $\pi$ phase shift [32]. Also, Figs. 9(g)-9(j) show temporal profiles and spectrogram plots [43] of RF up-chirp and down-chirp waveforms, respectively, each with 10 GHz center frequency, TBWP of 8, and chirp rate of 8 GHz/ns.

Another example of on-chip RF-AWG is demonstrated in [33]. The schematic for the fabricated on-chip pulse shaper is plotted in Fig. 10. Here, two linear arrays each comprising eight cascaded microrings of different radii are coupled together via thermally-controlled delay lines. Unlike the spectral shaping structure of Fig. 9(a), the principle of operation for this RF-AWG setup is based on direct time-domain waveform synthesis. Each microring in the top linear array spectrally samples an ultrashort optical input pulse, thereby producing 8 replica pulses made up of distinct wavelength components. The various replicas are passed through independently tunable delay lines before being recombined by the bottom row of microrings. Each delay line includes a 25 ps fixed delay increment implemented via different waveguide lengths and a tunable component that allows control exceeding the 25 ps fixed delay increment. For each delay line the tunable component is implemented using a bank of 41 all-pass microring filters which offer a thermally tunable, resonant delay response. Thermal fine tuning of the relative resonant frequencies of top and bottom microrings allows control over the amplitude of the various replica pulses. RF-AWG operation was demonstrated via generation of both burst and continuous 30 and 40 GHz waveforms, waveforms incorporating abrupt phase shifts, and fast chirp waveforms with frequency content ranging from 30 GHz to 50 GHz. Several hundred microrings contribute to provide such waveform generation functionality.

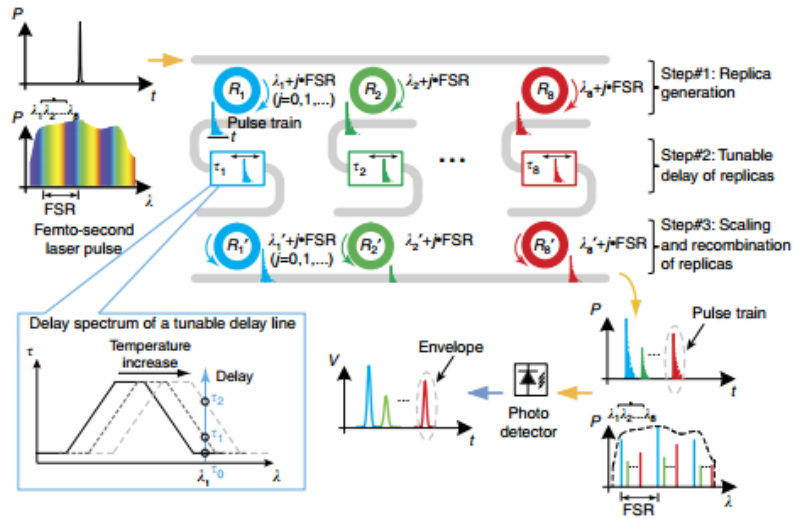

Fig. 10. Schematic of time-domain synthesis method where a femtosecond laser pulse is replicated, delayed and recombined through an eight-channel Si pulse shaper [33]. The inset schematically shows the mechanism of the thermally-tunable optical delay lines [33]. FSR: free spectral range of a microring.

### III. ULTRABROADBAND APPLICATIONS

Advances in the photonic synthesis of programmable ultrabroadband RF waveforms facilitate a wide range of ultrabroadband RF applications. In this section we discuss several recent proof-of-concept demonstrations of photonic-assisted RF-AWG. These include generation of a MMW Costas sequence (a favorable radar waveform), high-resolution unambiguous ranging in the W-band, and sounding and compensation of UWB multipath communication channels. These examples serve not only to illustrate applications possibilities, but also to demonstrate that photonic waveform generation capability is now sufficiently developed to support application studies.

*A. Millimeter-Wave Costas Sequence Generation*

With the freedom to arbitrarily tailor the synthesized waveform, it is natural to seek to generate waveforms other than the continuous frequency chirps shown in most of our previous examples. One example is the generation of MMW Costas sequences, a family of frequency-hopped spread spectrum waveforms which are widely used in applications such as radar engineering, synchronization, and communications [77]. A wideband MMW Costas sequence is characterized by a thumbtack auto-ambiguity function [77], indicating strong performance in both range resolution and Doppler. The generation of a wideband MMW Costas sequence was experimentally demonstrated utilizing the interferometric photonic RF-AWG scheme of [59]. As illustrated in Fig. 11(a), this Costas sequence has length 15, is centered around 10 GHz, and executes frequency hops to discrete frequencies arranged on a 1 GHz frequency grid. To evaluate this sequence, the auto-ambiguity function of the recorded waveform in computed offline and plotted in Fig. 11(b). The nice thumbtack shape of the auto-ambiguity function, with a full width at half maximum (FWHM) of ~15 ps in delay and ~36 MHz in Doppler, is in good agreement with expectations from the target sequence. As we discuss in Section IV, time-aperture expansion techniques exist that could substantially increase the relatively short temporal duration of the generated waveform to achieve finer Doppler resolution [78]. Later we successfully extended Costas sequence generation to the W-band (75-110 GHz) [60]. A down-conversion measurement of the generated waveform is shown in Fig. 11(c). In this experiment a different length-15 Costas sequence was implemented, again with hops to frequencies positioned along a 1-GHz-spaced grid but now with a center frequency of 90 GHz.

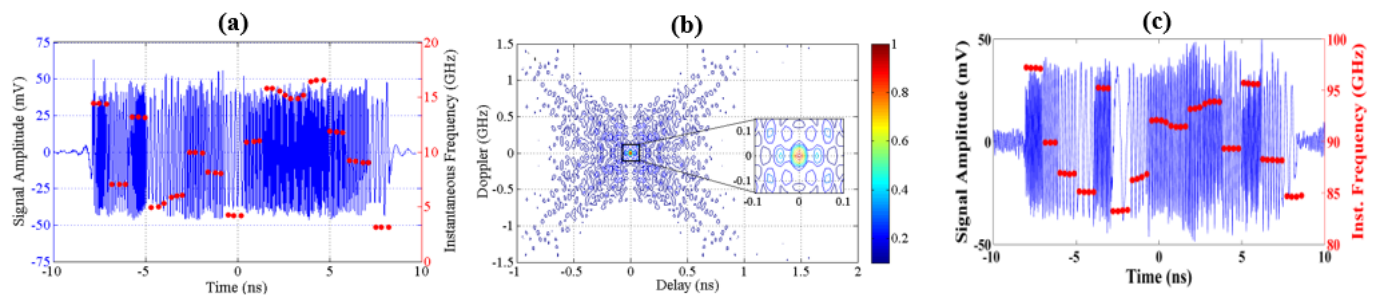

Fig. 11. Wideband MMW Costas sequence generation – (a) Time domain measurement of Costas sequence at 10 GHz center frequency and 1 GHz frequency increment (in blue) and corresponding instantaneous frequency (in red). (b) Normalized contour plot of auto-ambiguity function of measurement of Fig. 11(a) [59]. (c) Time-domain down-conversion measurement of W-band Costas sequence with 90 GHz center frequency and 1 GHz frequency increment (in blue) and corresponding instantaneous frequency (in red) [60].

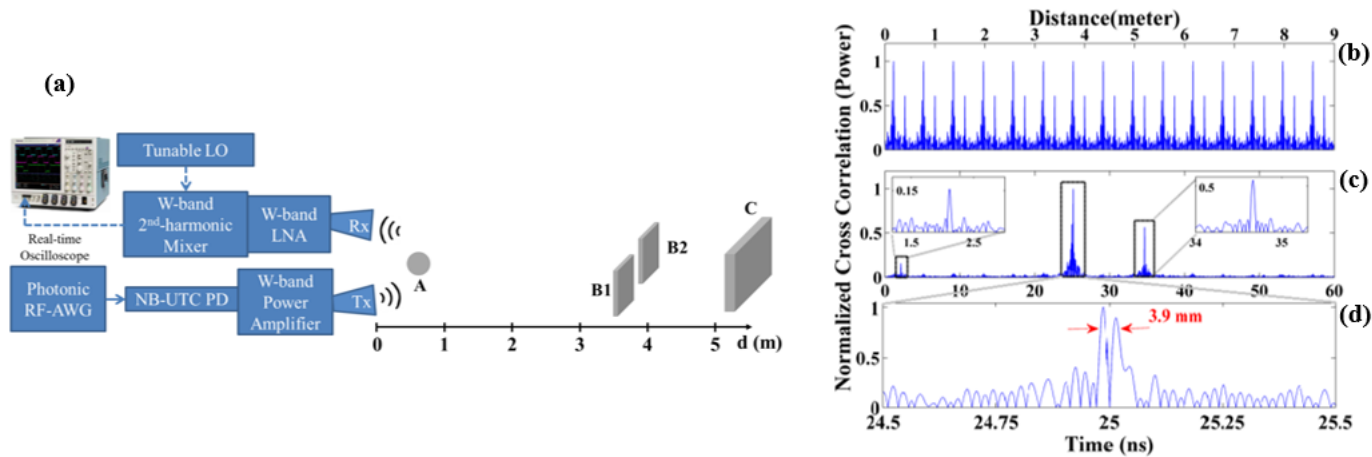

Fig. 12. Ultrahigh-resolution W-band ranging experiments [60] – (a) Setup schematic. Ball A with radius 4 cm. Flat sheets B1 and B2 are 10×10 cm. Flat sheet C is 10×15 cm. (b) W-band ranging without time aperture expansion. (c) W-band ranging with length-15 PN-modulated sensing waveforms. (d) Magnified view from 24.5–25.5 ns of Fig. 12(c), depicting range resolution. Tx: Transmitter, Rx: Receiver, LNA: Low-noise amplifier.

### B. High-Resolution Unambiguous Ranging in W-band

A multiple object ranging experiment in the W-band [60] offers another application example utilizing the broad bandwidth at high center frequencies offered by photonic RF-AWG. The use of high MMW frequency bands not only circumvents the congestion problem of lower frequency regions, but also provides large bandwidth with promise for high-speed communication and high-resolution resolution ranging [1], [4]–[6], [77]. The ranging setup is sketched in Fig. 12(a). Shaped optical pulses from the FTM-based photonic RF-AWG subsystem are directed to a near-ballistic uni-travelling carrier photodiode (NBUTC-PD) chip which offers good high frequency response [79]. This creates RF signals in the W-band which are amplified electronically and transmitted by a horn antenna. Echoes from the objects are captured by a second, receiving antenna, amplified, mixed with a local oscillator (LO) down to baseband, and recorded on a real-time oscilloscope. Four aluminum objects are placed at different distances to serve as targets. W-band linear frequency chirp waveforms with bandwidths up to 30 GHz and TBWP of ~600 are used as the probing signals. Offline cross-correlation is performed to compress the chirped return signals to the bandwidth limit and determine the ranges of the objects.

Figs. 12(b)-12(d) show examples of the results. The extremely broad bandwidth provided by photonic RF-AWG resulted in an extremely fine range resolution. Fig. 12(d) shows distinct cross-correlation peaks corresponding to aluminum plates B1 and B2 separated by only 3.9 mm. To our knowledge the achieved range resolution is significantly better than the previous best reported in the W-band (12mm) [6] and also improves, but to a lesser extent, on resolutions reported for experiments with signals generated electronically in considerably higher MMW and sub-THz frequency bands [4], [5]. However, an interesting point is that the 4 ns repetition period of the mode-locked laser used in photonic RF-AWG is much less than the range of delays associated with the return signals from different objects within the ~5 meter ranging environment. Therefore, multiple copies of each return signal are folded into ranging data of Fig. 12(b); weaker return signals are hidden within stronger ones and the ranges become ambiguous. By expanding the repeat-free time aperture of the sounding waveforms, the ambiguity is removed. Our technique for time aperture expansion is reported in [78] and is discussed in detail in Section IV of the current paper. Fig. 12(c) shows data obtained with the repetition period of the sounding waveform increased from 4 ns to 60 ns. The ranges of the objects in the environment are now unambiguously detected.

### C. Ultra-wideband Channel Sounding and Compensation

Due to the broad bandwidths involved, wireless transmission of UWB signals is highly sensitive to dispersion and multipath propagation. As a result of multipath, the transmitted signal arrives at the receiver through different paths with different delays and attenuations. These multiply delayed signal components add up constructively at some frequencies and destructively at others. Furthermore, the intrinsic dispersion of certain classes of broadband transmitter and receiver antennas also contributes significantly [80]–[82]. Consequently, for broadband RF propagation channels, compensation must be applied to compress received signals into short pulses and for high rate communications. In the RF propagation literature, such compression is termed temporal focusing. Interestingly, it has been shown that temporal focusing of broad bandwidth signals in dense multipath environments is usually accompanied by spatial focusing, meaning that the compensation applied to obtain pulse compression at a particular receiver remains effective only within a limited spatial region around that receiver. Such spatial focusing occurs because the multipath channel response decorrelates rapidly with position. These effects offer interesting opportunities for applications such as covert communications [83]–[85].

Ultrabroadband photonic-assisted RF-AWG can be utilized for characterization and compensation of antenna dispersion

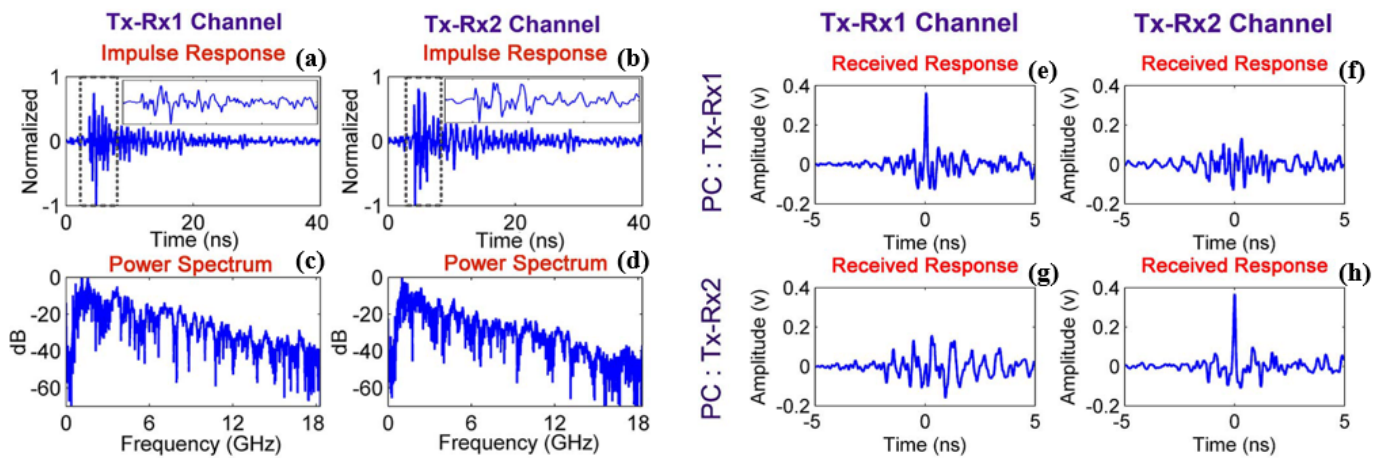

Fig. 13. Channel sounding and compensation of highly-dense multipath UWB communication channels using photonic RF-AWG [84] – (a), (b) Impulse responses of Tx–Rx1 and Tx–Rx2 channels, and (c), (d) Corresponding power spectra of the Tx–Rx1 and Tx–Rx2 links. (e), (f) When the phase compensated (PC) transmit waveform is designed based on the Tx–Rx1 response, a significant peaking results at the target receiver. Receiver Rx2 located just a couple of wavelengths away from the target receiver gets only a noise like interference. (g), (h) When the Tx transmits a waveform designed to achieve peaking at Rx2, the received signal by Rx1 now remains noise-like.

[57], [80]–[82] and dense UWB multipath [83]–[85]. To obtain the impulse response of the wireless channel (which contains multipath and dispersion information), spread spectrum channel sounding can be utilized [86]. Our group has previously published on characterization and compensation of UWB signals distorted by the frequency dependent delay inherent to certain broadband antennas [57], [80]–[82]. Here we focus on our investigations of multipath channels. Multipath channel sounding experiments using photonic RF-AWG for the frequency range 2-18 GHz are presented in [84]. There, RF chirps spanning 2-18 GHz are synthesized using the NF-FTM technique and excite the transmit antenna (Tx), which radiates into the multipath environment. The received waveform is recorded by a real-time oscilloscope, and a deconvolution algorithm [83], [86] is employed offline to retrieve the impulse response of the propagation channel. In the experiment a pair of receive antennas (Rx1 and Rx2, separated from each other by ~50 cm) are placed ~10 meters away from the transmit antenna with an intervening wall in a non-line-of-sight (NLOS) geometry. The impulse responses of the Tx-Rx1 and Tx-Rx2 links are plotted in Fig. 13(a) and 13(b), respectively. For both cases, complicated temporal structures spreading over 30 ns are detected with high resolution. Figs. 13(c) and 13(d) show the spectral representation of the channel, portraying a significant roll off with increase of frequency, as well as deep fades at various frequencies where multipath components interfere destructively. Although the Tx-Rx1 and Tx-Rx2 responses are qualitatively similar, the details are quite different, providing evidence for the spatial decorrelation mentioned above.

To enable high-speed communication of information, compensation of strong multipath distortion, such as that seen in Figs. 13(a)-(d), is required. Channel compensation may be performed either on the transmitter end, in the form of predistortion, or on the receiver end, in the form of a matched (or other) filter. For many wireless applications, the former is preferred as it enables a much simpler receiving structure. In this discussion we consider phase compensation [87], in which the frequency dependent phase of the channel impulse response is extracted; and a waveform having the opposite spectral phase is transmitted to achieve pulse compression. When a predistorted waveform is used to excite the channel, a compressed pulse with FWHM less than 70 ps arrives at receiver Rx1, as plotted in Fig. 13(e). Here the utilized predistorted waveform is calculated based on the impulse response of the Tx-Rx1 link. Moreover, although the predistorted waveform is transmitted omnidirectionally, only a noise-like waveform is captured at Rx2, which is located just a few wavelengths away from the intended receiver Rx1 (Fig. 13(f)). If instead the photonic RF-AWG is programmed to generate the predistorted waveform calculated from the impulse response of the Tx-Rx2 link, the situation is reversed: the compressed pulse shows up at Rx2 and only a noise-like signal is observed at Rx1, as illustrated in Figs. 13(g) and 13(h), respectively. These experimental results demonstrate clear spatio-temporal compression over an ultrabroadband dense multipath wireless channel. Recently, we have achieved a significant extension of this work, demonstrating data transmission at 250 Mbits/sec over a similar multipath channel [88]. These works show that photonic RF-AWG-based systems are capable of performing not only signal distribution, but also sophisticated functionalities such as channel measurement, channel compensation, and high-speed communication. With its inherent scalability to higher frequency and wider bandwidth, photonic RF-AWG is expected to offer even more advances in low probability-of-intercept ultra-high data rate wireless communications in the future.

## IV. TIME APERTURE EXPANSION

We have mentioned earlier that although the TBWP capability of photonic-assisted RF-AWG schemes has now reached a level appropriate for some spread spectrum applications, photonic RF-AWG techniques still suffer from the problem of limited repeat-free time apertures, typically in the low nanosecond range. This limitation inhibits the use of the generated waveforms in applications like long-range radar, Doppler radar, channel sounding and compensation of long communication channels, and more. Early examples of

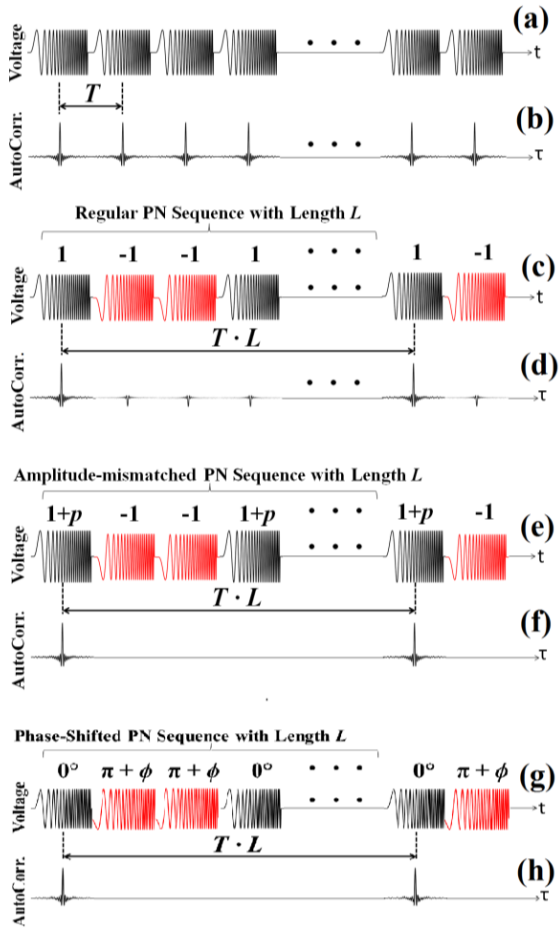

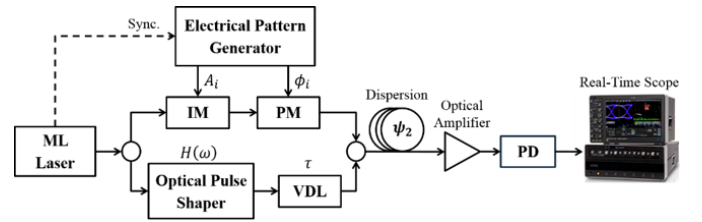

Fig. 15. Interferometric RF-AWG configuration with real-time phase and amplitude modulation capability [94].

Fig. 14. Schematic diagram of time aperture expansion RF-AWG, adapted from [78] – (a) Repetitive linear up-chirp waveform train with period $T$. (b) Autocorrelation function of (a). (c) Chirp train in (a) modulated by a length $L$ regular PN sequence. Polarity of chirps is either maintained (black) or flipped (red). (d) Autocorrelation function of (c), spacing of $T$ between low-power peaks. (e) Chirp train in (a) modulated by a length $L$ amplitude-mismatched PN sequence. Peak-to-peak amplitude of the polarity-maintained chirps in (c) is adjusted to $1+p$, where $p$ is a PN length dependent value. (f) Autocorrelation function of (e). (g) Chirp train in (a) modulated by a length $L$ phase-shifted PN sequence. An excess phase shift of $\phi$ is applied to antipodal waveforms. This value depends on the PN sequence length according to (8). (h) Autocorrelation function of (g).

RF-photonic waveform generation work trying to address this issue are reported in [63]–[66], where line-by-line optical pulse shaping accompanied by ultrafast switching is utilized to construct frequency-hopped microwave pulses with indefinite record lengths and controllable RF spectra. Recently, we reported a novel photonic spread spectrum radio-frequency waveform generation technique that combines FTM-based RF-AWG and pseudo-noise (PN) waveform switching, implemented in the optical domain using an electro-optic modulator-based switch. This increases the repeat-free time aperture of FTM-based RF-AWG essentially arbitrarily, up to several microseconds and beyond [78].

Fig. 14 shows the principle of operation. In the absence of PN coding, the output of the photonic RF-AWG system is comprised of programmable RF waveforms (linear frequency chirped waveform for instance, as shown in Fig. 14(a)), constrained to repeat at the fixed repetition rate of the utilized ML laser ($T$). Consequently, after pulse compression, achieved by autocorrelation, we observe short peaks occurring every $T$ seconds, Fig. 14(b). In our work so far, pulse compression has been implemented offline using post-processing of the measured waveforms. However, real-time compression using programmable wideband RF phase filters implemented using photonic signal processing is an option for future research [89]–[92]. When a length $L$ binary PN sequence [93] is modulated onto such repetitive waveform (Fig. 14(c)), its repetition period can be extended to $T \cdot L$, which is the repeat period of the applied PN sequence. From the autocorrelation perspective, large autocorrelation peaks are now observed only at multiples of $T \cdot L$, while still there exists some small peaks with negative power at the original locations, Fig. 14(d). In [78], we present a mathematical relation for a slight amplitude-detuning ($p = 2/\sqrt{L+1}$) of the applied PN sequence (the PN sequence now takes on values 1+p and -1, as opposed to 1 and -1 conventionally). This detuning factor allows the suppression of the residual unwanted peaks, achieving true expansion of the repeat-free time aperture, schematically seen in Figs. 14(e) and 14(f). In [78] we reported tunable time aperture expansion of ultrabroadband microwave signals with frequency content of up to 20 GHz, from 4 nanoseconds up to several microseconds.

Although the amplitude-detuned PN coding scheme of [78] do allow increased repetition periods, this scheme incurs some loss of power, especially when the length of the PN sequence is short. For example, to expand the repetition period by a factor of 15 results in a $p$ value of $1/2$, corresponding to ~30% average power decrease. This is a significant disadvantage for many applications, considering the peak-voltage limited nature of typical RF transmitters. To tackle this problem, here we propose time aperture expansion based on modulation of the basis waveforms by phase-detuned PN sequences. The principle of this technique is illustrated in Figs. 14(g) and 14(h). In addition to the polarity flipping sketched in Fig. 14(c), which is essentially a phase shift of $\pi$, a small phase detuning of $\Delta\phi$ is introduced. The phase-detuned PN sequence takes on phase values 0 and $\pi + \Delta\phi$, with constant amplitude. According to the mathematical derivation shown in the Appendix, the phase detuning should be set to:

$$\Delta\phi = \cos^{-1}\left(\frac{L-1}{L+1}\right) \quad (6)$$

where $L$ is the length of the PN modulation sequence. With this setting, residual unwanted autocorrelation peaks disappear, leaving only strong peaks separated by $T \cdot L$. This phase-detuned PN modulation scheme allows arbitrary expansion of the repetition period while preserving the average power of the output waveform under a peak-voltage-limited transmitter.

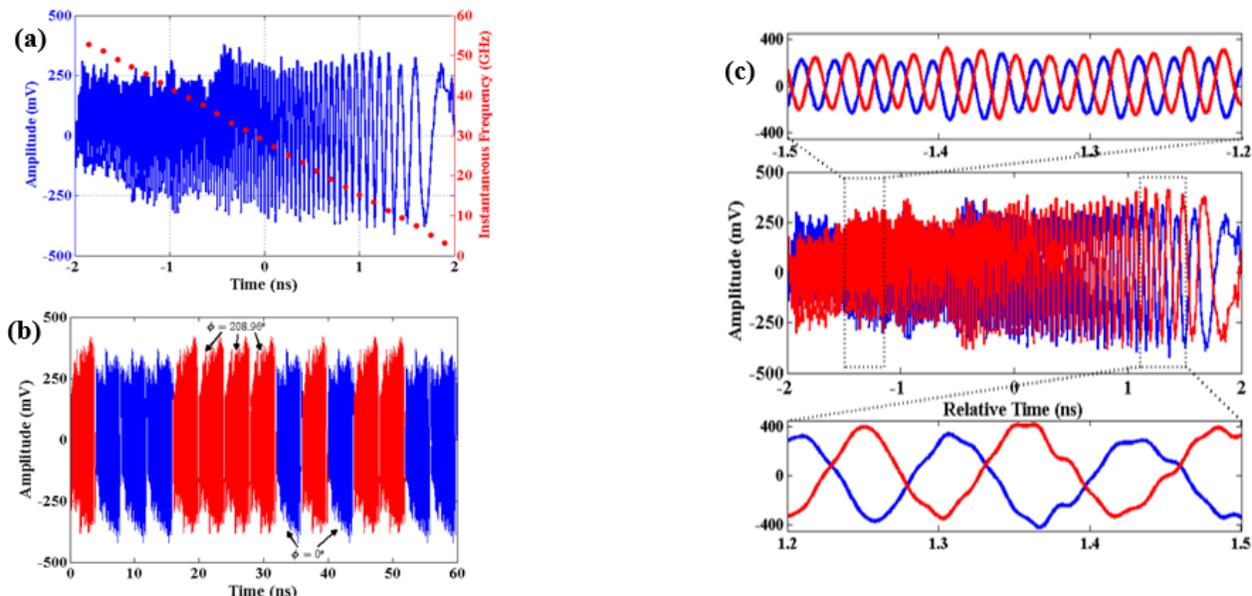

Fig. 16. Measurement results of ultrabroadband linear down-chirp spanning 2-52 GHz – (a) Single waveform and its instantaneous frequency plot. (b) Color-coded 60 ns window of 180 ns-long measurement of length-15 phase-detuned PN coding according to Eq. (15). (c) Overlay of two phase-coded basis waveforms of Fig. 16(b) and zoom-in views.

To demonstrate the proposed repetition period expansion in experiment, we recently introduced a modification of our interferometric RF-AWG setup, capable of real-time phase and amplitude modulation of the individual arbitrary waveforms in the generated sequence, independent of the RF waveform design stage [94]. The schematic diagram for this setup is illustrated in Fig. 15, comprised of two interferometer arms. The bottom arm is a replica of the original interferometric RF-AWG system, with an optical pulse shaper and variable delay line for passband RF-AWG functionality. The top arm is now provisioned with high-speed optical phase and intensity modulators (PM and IM), which are driven by an externally applied data sequence. For our purposes here, we only use the phase-coding capability of this setup.

In [94] we used a high-speed real-time oscilloscope to measure time aperture expanded ultrabroadband signals covering the 4.7 octave frequency range of 2-52 GHz. Fig. 16(a) shows a measured ultrabroadband downchirp waveform along with its computed instantaneous frequency plot. By applying a quadratic spectral phase function on the optical pulse shaper in the setup of Fig. 15, a linearly-decreasing frequency function is achieved across the center frequency of 27 GHz, which is set by tuning the VDL. In the absence of phase coding, this RF waveform repeats at the repetition rate of the ML laser (250 MHz), making the repeat-free time aperture 4 ns and the TBWP of the generated RF waveform 200 (50 GHz × 4 ns). To demonstrate time aperture expansion via the phase-detuned PN modulation technique, we program the pattern generator with a length 15 phase-detuned PN sequence and apply a phase detuning of $\Delta\phi = 28.96°$, as derived from (8). Fig. 16(b) shows a color-coded 60 ns window of a 180 ns-long measurement for this experiment. The corresponding phase values for each basis waveform are also indicated in the figure. To provide a clearer observation, the two phase-coded basis waveforms from this sequence are overlaid in Fig. 16(c) along with zoom-in views. The detuning from pure $\pi$ phase modulation is easily observed.

To evaluate temporal period expansion, we compute the circular autocorrelation of the 180 ns-long measurement as well as that of a same length measurement of RF-AWG in the absence of coding. In Fig. 17(a), the normalized autocorrelation of the unmodulated waveform has a period of 4 ns, the same as the laser repetition period. In contrast, as shown in Fig. 17(b), we have achieved an autocorrelation with strong peaks separated by 60 ns, corresponding to an increase in period by a factor of the PN sequence length (15 here). In the absence of phase detuning, we expect to observe weak autocorrelation peaks with a normalized amplitude of -1/15 occurring every 4 ns. Consequently, a peak extinction ratio of ~ -11.76 dB is anticipated. This value is defined as the ratio between the strongest amplitude of the remaining unwanted peaks and the strong peak at zero delay. However, as observed from the magnified view of Fig. 17(b), phase detuning suppresses these peaks down to smaller than 3e$^{-4}$, or equivalently ~ -35.23 dB. To show the potential of increasing the repeat-free time aperture even further, we extend the length of the PN sequence and take real-time measurements of the generated waveforms. Fig. 17(c) is the computed autocorrelation function of the generated chirped waveform with a length-1023 phase-detuned PN sequence modulation. In this case, the time aperture is increased to more than 4 microseconds.

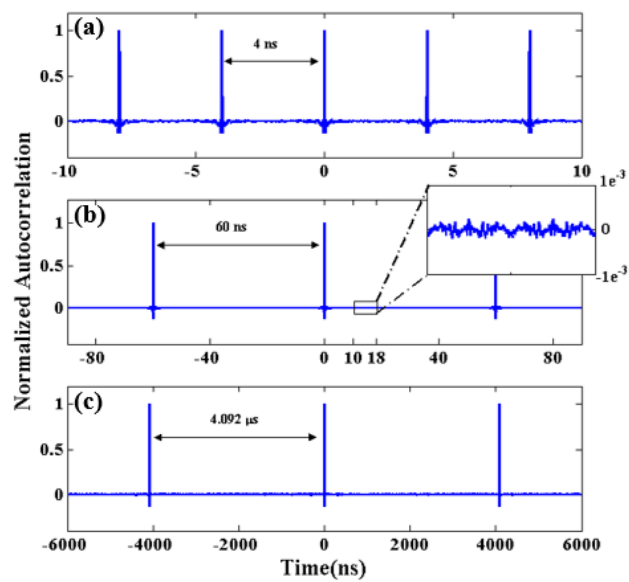

Fig. 17. Normalized autocorrelation functions of generated ultrabroadband chirp sequences. (a) No Modulation, (b) Length 15 phase-detuned PN modulation, (c) Length 1023 phase-detuned PN modulation.

Finally, it is worth noting that from the frequency domain perspective, time aperture expansion is equivalent to having a frequency spectrum with finer spaced modes, which can be utilized for high-resolution channel sounding experiments as well as Doppler radar. For the examples of Fig. 17, the generated RF-AWG power spectra have frequency resolution of 250 MHz, ~16.67 MHz and ~244.38 KHz, respectively.

## V. Discussions and Conclusion

In this paper we have reviewed programmable photonic-assisted RF arbitrary waveform generation, with an emphasis on recent advances leading to individual ultrabroadband RF waveforms with TBWPs up to ~600. With the addition of waveform by waveform pseudorandom phase keying, further expansion of time aperture and TBWP is possible, fundamentally without limit. Such waveform generation techniques would generally be considered as transmitter technologies, as they are used to provide arbitrary waveforms at a wireless or radar transmitter. Photonic signal processing may also be applied advantageously as a receiver technology, i.e., for real-time processing of radar return signals or incoming communications waveforms. In the context of systems employing ultrabroadband arbitrary waveform transmission, photonically implemented phase filters offer a potentially attractive solution for realizing pulse compression functionality in the receiver. Early RF photonic phase filters based on hyperfine resolution optical pulse shapers have been demonstrated for compression of photonically generated RF waveforms with TBWPs of order ten but without wireless transmission [89]. Newer RF photonic phase filters based on dispersive propagation of optical frequency comb sources are capable of larger TBWP and have recently been demonstrated for compression of electronically generated spread spectrums signals [91], [92]. In the future we anticipate systems in which photonics will be used both for generation and wireless transmission of large TBWP radio frequency signals and for subsequent pulse compression processing in the receiver.

## Appendix

In photonic-assisted RF-AWG, a repeating optical pulse train from a pulsed laser source is fed into the generation setup, resulting in arbitrarily-tailored RF waveforms that repeat periodically at the laser repetition rate. This RF waveform sequence, $s(t)$, is expressed as:

$$s(t) = \sum_i w(t - iT) \tag{A.1}$$

where $w(t)$ is the temporal profile of each spread spectrum waveform with circular autocorrelation function $R_w(\tau)$, and $T$ is the repetition period. As a result, the autocorrelation function of $s(t)$, $R_s(\tau)$, equals:

$$R_s(\tau) = \int s(t) \cdot s(t + \tau) \, dt = \sum_i R_w(\tau - iT) \tag{A.2}$$

We must note that for the continuous signal, $w(t)$, the autocorrelation is defined as below:

$$R_w(\tau) = \int w(t) \cdot w(t + \tau) dt \tag{A.3}$$

Clearly from (A.2), we observe that in the absence of coding, due to the fact that $w(t)$ is repeating itself with period $T$, the autocorrelation function of the total waveform is just a repetitive sum of the autocorrelation of a single waveform, $w(t)$. However, we may use simple coding schemes on each waveform to suppress these repeating autocorrelation peaks for a longer duration and combine the power into the main autocorrelation peak at zero delay [78]. Binary ($\pm 1$) PN sequences of length $L = 2^m - 1$ [93], have a thumb-tack like circular (or periodic) autocorrelation, i.e.:

$$r_{PN}[m] = \sum_{i=1}^{L} PN[i] \cdot \overline{PN}[(i + m) \bmod L]$$
$$= \begin{cases} L & m = 0, \pm L, \pm 2L, \ldots \\ -1 & \text{elsewhere} \end{cases} \tag{A.4}$$

where $PN[i]$ is the binary pseudorandom sequence of repetition length $L$, and $r_{PN}[m]$ is its discrete circular autocorrelation function. The bar sign in $\overline{PN}$ denotes the complex conjugate function.

In order to modulate this sequence on the repetitive pulse sequence of (A.1), we must phase shift certain pulses by $\pi$ according to $PN[i]$. However, that would result in a non-zero correlation floor with a value of -1, as seen in (A.4). In the following derivations, we prove that by detuning the phase shift slightly from $\pi$, we can achieve perfect autocorrelation with zero correlation floor. This modified phase-detuned PN sequence is referred to as $PN'[i]$ and is composed as below:

$$PN'[i] = \frac{1 + e^{j\Delta\phi}}{2} PN[i] + \frac{1 - e^{j\Delta\phi}}{2} U[i] \tag{A.5}$$

where $\Delta\phi$ is the constant phase detuning value in radians and $U$ is a unit sequence of length $L$ defined as $U[i] = 1$, for all $1 \leq i \leq L$. As a result, $PN'[i]$ has the same pattern as $PN[i]$, but instead of having a phase shift of $\pi$ at certain delay values, it exhibits phase shifts of $\pi + \Delta\phi$, making a complex-valued sequence. After some mathematical calculations, the discrete circular autocorrelation of $PN'[i]$ can be expressed as:

$$r_{PN'}[m] = \frac{1+\cos\Delta\phi}{2}r_{PN}[m] + \frac{1-\cos\Delta\phi}{2}r_U[m]$$
$$+ \frac{j\sin\Delta\phi}{2}r_{PN,U}[m] - \frac{j\sin\Delta\phi}{2}r_{U,PN}[m] \quad (A.6)$$

where $r_{PN}[m]$ and $r_U[m]$ are the discrete circular autocorrelation functions of $PN[i]$ and $U[i]$, while $r_{PN,U}[m]$ and $r_{U,PN}[m]$ are the discrete circular cross-correlations between the two sequences. Since both $PN[i]$ and $U[i]$ are real-valued sequences, the last two terms in (A.6) cancel out. Also, $r_U[m]$ is equal to $L$ for all values of delay, thus plugging in (A.4) into (A.6) results in

$$r_{PN'}[m] = \begin{cases} L & m = 0, \pm L, \pm 2L, \ldots \\ \frac{1}{2}(L-1) - \frac{1}{2}(L+1)\cos\Delta\phi & \text{elsewhere} \end{cases} \quad (A.7)$$

Clearly, by appropriately choosing the phase detuning parameter, we are now able to create a zero floor at unwanted delays in (A.7). This is achieved at $\Delta\phi$ given as follows:

$$\Delta\phi = \cos^{-1}\left(\frac{L-1}{L+1}\right) \quad (A.8)$$

If this sequence, $PN'[i]$, is utilized to modulate the phase of the generated spread spectrum pulse sequence, we can completely null the unwanted peaks, thereby achieving perfect time aperture expansion. The autocorrelation function of the new waveform, $s'(t)$, can now be expressed as

$$R_{s'}(\tau) = \sum_i L \cdot R_w(\tau - iLT) \quad (A.9)$$

Equation (A.9) indicates that the modulation perfectly increases the temporal repeat-free period of the new waveform by a factor of $L$. Also, the temporal profile of the basis waveform, $w(t)$, i.e., the temporal profile within a single period $T$, is not disturbed, leaving the RF bandwidth preserved. As a result, the spread spectrum waveform generation which controls the shape and RF bandwidth, and the PN modulation which determines the repeat-free time aperture, are kept independent while arbitrarily large TBWPs are realized.

REFERENCES

[1] A.Y. Nashashibi, K. Sarabandi, P. Frantzis, R.D. De Roo, F.T. Ulaby, "An Ultrafast Wide-Band Millimeter-Wave (MMW) Polarimetric Radar for Remote Sensing Applications," *IEEE T. Geoscience and Remote Sensing*, vol.40, no.8, pp.1777-1786, 2002.
[2] G.G. Brown, B.C. Dian, K.O. Douglass, S.M. Geyer, S.T. Shipman, B.H. Pate, "A Broadband Fourier Transform Microwave Spectrometer based on Chirped Pulse Excitation," *Rev. Sci. Instrum.*, vol.79, no.5, pp.053103-053103.13, 2008.
[3] C.W. Chow, L.G. Yang, C.H. Yeh, C.B. Huang, J.W. Shi, C.L. Pan, "10 Gb/s Optical Carrier Distributed Network with W-band (0.1THz) Short-Reach Wireless Communication System," *Optics Communications*, vol.285, pp.4307-4311, 2012.
[4] K.B. Cooper, R.J. Dengler, N. Llombart, T. Bryllert, G. Chattopadhyay, E. Schlecht, J. Gill, C. Lee, A. Skalare, I. Mehdi, and P.H. Siegel, "Penetrating 3-D Imaging at 4-and 25-m Range Using a Submillimeter-Wave Radar," *IEEE T. Microw. Theory Tech.*, vol.56, no.12, pp.2771-2778, 2008.
[5] K.B. Cooper, R.J. Dengler, N. Llombart, A. Talukder, A.V. Panangadan, C.S. Peay, I. Mehdi, and P.H. Siegel, "Fast, High-Resolution Terahertz Radar Imaging at 25 Meters," in *Proc. SPIE*, 2010, vol.7671, p.76710Y.
[6] T.F. Tseng, J.M. Wun, W. Chen, S.W. Peng, J.W. Shi, and C.K. Sun, "High-Depth-Resolution 3-Dimensional Radar-Imaging System based on a Few-Cycle W-band Photonic Millimeter-wave Pulse Generator," *Opt. Express*, vol.21, no.12, pp.14109-14119, 2013.
[7] M. Miyakawa and J.C. Bolomey, "Microwave Imaging I: Microwave Computed Tomography," in *Non-Invasive Thermometry of the Human Body*: CRC Press, 1995, pp.105-126.
[8] J.A. Zeitler and L.F. Gladden, "In-Vitro Tomography and Nondestructive Imaging at Depth of Pharmaceutical Solid Dosage Forms," *Eur. J. Pharmaceutics Biopharmaceutics*, vol.71, no.1, pp.2-22, 2009.
[9] D.M. Sheen, J.L. Fernandes, J.R. Tedeschi, D.L. McMakin, A.M. Jones, W.M. Lechelt, and R.H. Severtsen, "Wide-Bandwidth, Wide-Beamwidth, High-Resolution, Millimeter-wave Imaging for Concealed Weapon Detection", in *Proc. SPIE*, 2013, vol.8715, p.871509.
[10] H.J. Hansen, "Standoff Detection using Millimeter and Submillimeter Wave Spectroscopy," in *Proc. IEEE*, 2007, vol.95, pp.1691-1704.
[11] L. Chao, M.N. Afsar, and K.A. Korolev, "Millimeter Wave Dielectric Spectroscopy and Breast Cancer Imaging," in *7th European Microwave Integrated Circuits Conference (EuMIC)*, 2012, pp.572-575.
[12] A.P. Colombo, Y. Zhou, K. Prozument, S.L. Coy, and R.W. Field, "Chirped-Pulse Millimeter-wave Spectroscopy: Spectrum, Dynamics, and Manipulation of Rydberg-Rydberg Transitions," *J. Chem. Phys.* vol.138, p.014301, 2013.
[13] J.L. Neill, B.J. Harris, A.L. Steber, K.O. Douglass, D.F. Plusquellic, and B.H. Pate, "Segmented Chirped-Pulse Fourier Transform Submillimeter Spectroscopy for Broadband Gas Analysis," *Opt. Express*, vol.21, no.17, pp.19743-19749, 2013.
[14] J. Federici and L. Moeller, "Review of Terahertz and Subterahertz Wireless Communications," *J. Appl. Phys.*, vol.107, p.111101, 2010.
[15] T. Kleine-Ostmann and T. Nagatsuma, "A Review on Terahertz Communications Research," *J. Infrared, Millim. Terahertz Waves*, vol.32, no.2, pp.143-171, 2011.
[16] A.J. Seeds and K. Williams, "Microwave Photonics," *IEEE J. Lightw. Technol.*, vol.24, no.12, pp.4628-4641, 2006.
[17] J. Capmany and D. Novak, "Microwave Photonics Combines Two Worlds," *Nature Photonics*, vol.1, no.6, pp.319-330, 2007.
[18] J.P. Yao, "Microwave Photonics," *IEEE J. Lightw. Technol.*, vol.27, no.3, pp.314-335, 2009.
[19] S. Hardy, "Keysight Technologies Offers 65-GSa/s, 20-GHz Arbitrary Waveform Generator," *Lightwave Online*, vol.31, no.5, 2014.
[20] L.A. Samoska, "An Overview of Solid-State Integrated Circuit Amplifiers in the Submillimeter-wave and THz Regime," *IEEE T. Terahertz Sci. Technol.*, vol.1, no.1, pp.9-24, 2011.
[21] A. Kanno and T. Kawanishi, "Broadband Frequency-Modulated Continuous-wave Signal Generation by Optical Modulation Technique," *IEEE J. Lightw. Technol.*, vol.32, no.20, pp.3567-3572, 2014.
[22] J.P. Yao, "Photonic Generation of Microwave Arbitrary Waveforms," *Optics Communications*, vol.284, no.15, pp.3723-3736, 2011.
[23] V. Torres-Company and A. M. Weiner, "Optical Frequency Comb Technology for Ultra-Broadband Radio-Frequency Photonics," *Laser & Photonics Reviews*, vol.8, no.3, pp.368-393, 2013.
[24] T. Nagatsuma and Y. Kado, "Microwave Photonic Devices and Their Applications to Communications and Measurements," *PIERS Online*, vol.4, no.3, pp.376-380, 2008.
[25] A.J. Seeds, M.J. Fice, K. Balakier, M. Natrella, O. Mitrofanov, M. Lamponi, M. Chtioui, F. van Dijk, M. Pepper, G. Aeppli, A.G. Davies, P. Dean, E. Linfield, and C. Renaud, "Coherent Terahertz Photonics," *Opt. Express*, vol.21, no.19, pp.22988-23000, 2013.
[26] J.D. McKinney and A.M. Weiner, "Photonic Synthesis of Ultrabroadband Arbitrary Electromagnetic Waveforms," in *Microwave Photonics*, 2nd ed., C.H. Lee Ed. USA: CRC Press, 2013, pp.243-306.
[27] S.T. Cundiff and A.M. Weiner, "Optical Arbitrary Waveform Generation," *Nat. Photonics*, vol. 4, no.11, pp.760-767, 2010.
[28] M. Li, J. Azaña, N. Zhu, and J.P. Yao, "Recent Progresses on Optical Arbitrary Waveform Generation," *Frontiers of Optoelectronics*, vol.7, no.3, pp.359-375, 2014.
[29] L.R. Chen, "Photonic Generation of Chirped Microwave and Millimeter Wave Pulses based on Optical Spectral Shaping and Wavelength-to-Time Mapping in Silicon Photonics," *Optics Communications*, in press, 2015. DOI: 10.1016/j.optcom.2015.04.023
[30] F.M. Soares, N.K. Fontaine, R.P. Scott, J.H. Baek, X. Zhou, T. Su, S. Cheung, Y. Wang, C. Junesand, S. Lourdudoss, K.Y. Liou, R.A. Hamm, W. Wang, B. Patel, L.A. Gruezke, W.T. Tsang, J.P. Heritage, S.J.B. Yoo, "Monolithic InP 100-Channel ×10-GHz Device for Optical Arbitrary Waveform Generation," *IEEE Photonics J.*, vol.3, no.6, pp.975-985,2011.

[31] C.G.H. Roeloffzen, L. Zhuang, C. Taddei, A. Leinse, R.G. Heideman, P.W.L. van Dijk, R.M. Oldenbeuving, D.A.I. Marpaung, M. Burla, K.J. Boller, "Silicon Nitride Microwave Photonic Circuits," *Opt. Express*, vol.21, no.19, pp.22937-22961, 2013.

[32] M.H. Khan, H. Shen, Y. Xuan, L. Zhao, S.J. Xiao, D.E. Leaird, A.M. Weiner, and M. Qi "Ultrabroad-Bandwidth Arbitrary Radiofrequency Waveform Generation with a Silicon Photonic Chip-based Spectral Shaper," *Nat. Photonics*, vol.4, no.2, pp.117-U130, 2010.

[33] J. Wang, H. Shen, L. Fan, R. Wu, B. Niu, L.T. Varghese, Y. Xuan, D.E. Leaird, X. Wang, F. Gan, A.M. Weiner, and M. Qi, "Reconfigurable Radio-Frequency Arbitrary Waveforms Synthesized in a Silicon Photonic Chip," *Nature Communications*, vol.6, 5957, 2015. DOI: 10.1038/ncomms6957

[34] W. Zhang, W. Liu, W. Li, H. Shahoei, and J.P. Yao, "Independently Tunable Multichannel Fractional-Order Temporal Differentiator Based on a Silicon-Photonic Symmetric Mach-Zehnder Interferometer Incorporating Cascaded Microring Resonators," *IEEE J. Lightw. Technol.*, vol.33, no.2, pp.361-367, 2015.

[35] B. Guan, S.S. Djordjevic, N.K. Fontaine, L. Zhou, S. Ibrahim, R.P. Scott, D.J. Geisler, Z. Ding, and S.J.B. Yoo, "CMOS Compatible Reconfigurable Silicon Photonic Lattice Filters using Cascaded Unit Cells for RF-Photonic Processing," *IEEE J. Sel. Top. Quantum Electron.*, vol.20, no.4, 8202110, 2014.

[36] D.E. Leaird and A.M. Weiner, "Femtosecond Direct Space-to-Time Pulse Shaping," *IEEE J. Quantum Electron.*, vol.32, no.4, pp.494-504, 2001.

[37] D.E. Leaird and A.M. Weiner, "Femtosecond Direct Space-to-Time Pulse Shaping in an Integrated-Optic Configuration," *Opt. Lett.*, vol.29, no.13, pp.1551-1553, 2004.

[38] J.D. McKinney, D.E. Leaird, and A.M. Weiner, "Millimeter-Wave Arbitrary Waveform Generation with a Direct Space-to-Time Pulse Shaper," *Opt. Lett.*, vol.27, no.15, pp.1345-1347, 2002.

[39] J.D. McKinney, D. Seo, D.E. Leaird, and A.M. Weiner, "Photonically Assisted Generation of Arbitrary Millimeter-wave and Microwave Electromagnetic Waveforms via Direct Space-to-Time Optical Pulse Shaping," *IEEE J. Lightw. Technol.*, vol.21, no.12, pp.3020-3028, 2003.

[40] R. Ashrafi M. Li and J. Azana, "Multi-TBaud Optical Coding based on Superluminal Space-to-Time Mapping in Long Period Gratings," *Scientific Research*, vol.3, no.2B, pp.126-130, 2013.

[41] R. Ashrafi, M. Li, N. Belhadj, M. Dastmalchi, S. LaRochelle, and J. Azaña, "Experimental Demonstration of Superluminal Space-to-Time Mapping in Long Period Gratings," *Opt. Lett.*, vol.38, no.9, pp.1419-1421, 2013.

[42] A.M. Weiner, "Femtosecond Pulse Shaping using Spatial Light Modulators," *Rev. Sci. Instrum.*, vol.71, no.5, pp.1929-1960, 2000.

[43] A.M. Weiner, *Ultrafast Optics*: Wiley, 2009.

[44] A.M. Weiner, "Ultrafast Optical Pulse Shaping: A Tutorial Review," *Optics Communications*, vol.284, no.15, pp.3679-3692, 2011.

[45] J. Chou, Y. Han, and B. Jalali, "Adaptive RF-photonic Arbitrary Waveform Generator," *IEEE Photon. Technol. Lett.*, vol.15, no.4, pp.581-583, 2003.

[46] I. Lin, J.D. McKinney, and A.M. Weiner, "Photonic Synthesis of Broadband Microwave Arbitrary Waveforms Applicable to Ultra-Wideband Communication," *IEEE Microw. Wireless Compon. Lett.*, vol.15, no.4, pp.226- 228, 2005.

[47] J.D. McKinney, I.S. Lin, and A.M. Weiner, "Shaping the Power Spectrum of Ultra-Wideband Radio-Frequency Signals," *IEEE T. Microw. Theory Tech.*, vol.54, no.12, pp.4247-4255, 2006.

[48] V. Torres-Company, J. Lancis, P. Andrés, and L.R. Chen, "20 GHz Arbitrary Radio-Frequency Waveform Generator based on Incoherent Pulse Shaping," *Opt. Express*, vol.16, no.20, pp.21564-21569, 2008.

[49] A. Zeitouny, S. Stepanov, O. Levinson, and M. Horowitz, "Optical Generation of Linearly Chirped Microwave Pulses using Fiber Bragg Gratings," *IEEE Photon. Technol. Lett.*, vol.17, no.3, pp.670-672, 2005.

[50] J. Azana, N. K. Berger, B. Levit, and B. Fischer, "Broadband Arbitrary Waveform Generation based on Microwave Frequency Upshifting in Optical Fibers," *IEEE J. Lightw. Technol.*, vol.24, no.7, pp.2673-2675, 2006.

[51] Y. Dai, X. Chen, H. Ji, and S. Xie, "Optical Arbitrary Waveform Generation based on Sampled Fiber Bragg Gratings," *IEEE Photon. Technol. Lett.*, vol.19, no.23, pp.1916-1918, 2007.

[52] W. Chao and J.P. Yao, "Photonic Generation of Chirped Millimeter-Wave Pulses Based on Nonlinear Frequency-to-Time Mapping in a Nonlinearly Chirped Fiber Bragg Grating," *IEEE T. Microw. Theory Tech.*, vol.56, no.2, pp.542-553, 2008.

[53] C. Wang and J. P. Yao, "Large Time-Bandwidth Product Microwave Arbitrary Waveform Generation using a Spatially Discrete Chirped Fiber Bragg Grating," *IEEE J. Lightw. Technol.*, vol.28, no.11, pp.1652-1670, 2010.

[54] M. Li and J.P. Yao, "Photonic Generation of Continuously Tunable Chirped Microwave Waveforms based on a Temporal Interferometer Incorporating an Optically-Pumped Linearly-Chirped Fiber Bragg Grating," *IEEE T. Microw. Theory Technol.*, vol.59, no.12, pp.3531-3537, 2011.

[55] V. Torres-Company, D.E. Leaird, and A.M. Weiner, "Dispersion Requirements in Coherent Frequency-to-Time Mapping," *Opt. Express*, vol.19, no.24, pp.24718-24729, 2011.

[56] A. Dezfooliyan and A.M. Weiner, "Photonic Synthesis of High Fidelity Microwave Arbitrary Waveforms using Near Field Frequency to Time Mapping," *Opt. Express*, vol.21, no.19, pp.22974-22987, 2013.

[57] A. M. Weiner, A. Dezfooliyan, Y. Li, and A. Rashidinejad, "Selected Advances in Photonic Ultrabroadband Radio-Frequency Arbitrary Waveform Generation," in *International Topical Meeting on Microwave Photonics (MWP)*, pp.325-328, 2013.

[58] H.B. Gao, C. Lei, M.H. Chen, F.J. Xing, H.W. Chen, and S.Z. Xie, "A Simple Photonic Generation of Linearly Chirped Microwave Pulse with Large Time-Bandwidth Product and High Compression Ratio," *Opt. Express*, vol.21, no.20, pp.23107-23115, 2013.

[59] A. Rashidinejad and A.M. Weiner, "Photonic Radio-Frequency Arbitrary Waveform Generation with Maximal Time-Bandwidth Product Capability," *IEEE J. Lightw. Technol.*, vol.32, no.20, pp.3383-3393, 2014.

[60] Y. Li, A. Rashidinejad, J.M. Wun, D.E. Leaird, J.W. Shi, and A.M. Weiner, "Photonic Generation of W-band Arbitrary Waveforms with High Time-Bandwidth Products Enabling 3.9mm Range Resolution," *Optica*, vol.1, no.6, pp.446-454, 2014.

[61] J.W. Shi, F.M. Kuo, N.W. Chen, S.Y. Set, C.B. Huang, J.E. Bowers, "Photonic Generation and Wireless Transmission of Linearly/Nonlinearly Continuously Tunable Chirped Millimeter-Wave Waveforms With High Time-Bandwidth Product at W-Band," *IEEE Photonics J.*, vol.4, no.1, pp.215-223, 2012.

[62] W. Liu and J.P. Yao, "Photonic Generation of Microwave Waveforms Based on a Polarization Modulator in a Sagnac Loop," *IEEE J. Lightw. Technol.*, vol.32, no.20, pp.3637-3644, 2014.

[63] C.B. Huang, D.E. Leaird and A.M. Weiner, "Time-Multiplexed Photonically Enabled Radio-Frequency Arbitrary Waveform Generation with 100ps Transitions," *Opt. Lett.*, vol.32, no.22, pp.3242-3244, 2007.

[64] C.B. Huang, D.E. Leaird and A.M. Weiner, "Synthesis of Millimeter-Wave Power Spectra using Time-Multiplexed Optical Pulse Shaping," *IEEE Photon. Technol. Lett.*, vol.21, no.18, pp.1287-1289, 2009.

[65] C. M. Long, D. E. Leaird, and A. M. Weiner, "Photonically Enabled Agile RF Waveform Generation by Optical Comb Shifting," *Opt. Lett.*, vol.35, no.23, pp.3892-3894, 2010.

[66] V. Torres-Company, A.J. Metcalf, D.E. Leaird, and A.M. Weiner, "Multichannel Radio-Frequency Arbitrary Waveform Generation Based on Multiwavelength Comb Switching and 2-D Line-by-Line Pulse Shaping," *IEEE Photon. Technol. Lett.*, vol.24, no.11, pp.891-893, 2012.

[67] W. Li and J.P. Yao, "Generation of Linearly Chirped Microwave Waveform with an Increased Time-Bandwidth Product based on a Tunable Optoelectronic Oscillator and a Recirculating Phase Modulation Loop," *IEEE J. Lightw. Technol.*, vol.32, no.20, pp.3573-3579, 2014.

[68] M. Li, C. Wang, W. Li, J.P. Yao "An Unbalanced Temporal Pulse Shaping System for Chirped Microwave Waveform Generation," *IEEE T. Microw. Theory Tech.*, vol.58, no.11, pp.2968-2975, 2010.

[69] G. Baxter, S. Frisken, D. Abakoumov, Z. Hao, I. Clarke, A. Bartos, S. Poole, "Highly Programmable Wavelength Selective Switch based on Liquid Crystal on Silicon Switching Elements," in *Optical Fiber Communication Conference, 2006 and the 2006 National Fiber Optic Engineers Conference. OFC 2006*, pp.3.

[70] M.A. Roelens, J.A. Bolger, D. Williams, and B.J. Eggleton, "Multi-Wavelength Synchronous Pulse Burst Generation with a Wavelength Selective Switch," *Opt. Express*, vol.16, no.14, pp.10152-10157, 2008.

[71] Y.Q. Liu, S.G. Park, and A.M. Weiner, "Enhancement of Narrowband Terahertz Radiation from Photoconducting Antennas by Optical Pulse Shaping," *Opt. Lett.*, vol.21, no.21, pp.1762-1764, 1996.

[72] S.A. Diddams, S.M. Kirchner, T. Fortier, D. Braje, A.M. Weiner, and L. Hollberg, "Improved Signal-to-Noise Ratio of 10 GHz Microwave

Signals Generated with a Mode-filtered Femtosecond Laser Frequency Comb," *Opt. Express*, vol.17, no.5, pp.3331-3340, 2009.

[73] N.K. Fontaine, R.P. Scott, J. Cao, A. Karalar, W. Jiang, K. Okamoto, J.P. Heritage, B.H. Kolner, and S.J.B. Yoo, "32 Phase×32 Amplitude Optical Arbitrary Waveform Generation," *Opt. Lett.*, vol.32, no.7, pp.865-867, 2007.

[74] J.D. Taylor, *Introduction to Ultra-Wideband Radar Systems*: CRC Press, 1995.

[75] M. Ghavami, L.B. Michael and R. Kohno, *Ultra Wideband Signals and Systems in Communication Engineering*: Wiley, 2007.

[76] J.D. McKinney, "Background-Free Arbitrary Waveform Generation via Polarization Pulse Shaping," *IEEE Photon. Technol. Lett.*, vol.22, no.16, pp.1193-1195, 2010.

[77] N. Levanon and E. Mozeson, *Radar Signals*: Wiley-IEEE Press, 2004.

[78] Y. Li, A. Dezfooliyan, A.M. Weiner, "Photonic Synthesis of Spread Spectrum Radio Frequency Waveforms with Arbitrarily Long Time Apertures," *IEEE J. Lightw. Technol.*, vol.32, no.20, pp.3580-3587, 2014.

[79] J.W. Shi, F.M. Kuo, and J.E. Bowers, "Design and Analysis of Ultra-High-Speed Near-Ballistic Uni-Traveling-Carrier Photodiodes under a 50-$\Omega$ Load for High-Power Performance," *IEEE Photon. Tech. Lett.*, vol.24, no.7, pp.533-535, 2012.

[80] J.D. McKinney, D. Peroulis, and A.M. Weiner, "Dispersion Limitations of Ultra-Wideband Wireless Links and their Compensation via Photonically Enabled Arbitrary Waveform Generation," *IEEE T. Microw. Theory Tech.*, vol.56, no.3, pp.710-719, 2008.

[81] J.D. McKinney and A.M. Weiner, "Compensation of the Effects of Antenna Dispersion on UWB Waveforms via Optical Pulse-Shaping Techniques," *IEEE T. Microw. Theory Tech.*, vol.54, no.4, pp.1681-1686, 2006.

[82] J.D. McKinney, D. Peroulis, and A.M. Weiner, "Time-Domain Measurement of the Frequency-Dependent Delay of Broadband Antennas," *IEEE T. Ant. Propagat.*, vol.56, no.1, pp.39-47, 2008.

[83] A. Dezfooliyan and A.M. Weiner, "Evaluation of Time Domain Propagation Measurements of UWB Systems Using Spread Spectrum Channel Sounding," *IEEE T. Ant. Propagat.*, vol.60, no.10, pp.4855-4865, 2012.

[84] A. Dezfooliyan and A.M. Weiner, "Microwave Photonics for Space-Time Compression of Ultrabroadband Signals through Multipath Wireless Channels," *Opt. Lett.*, vol.38, no.23, pp.4946-4949, 2013.

[85] I.H. Naqvi, P. Besnier, G.E. Zein, "Robustness of a Time-Reversal Ultra-Wideband System in Non-Stationary Channel Environments," *IET Microw. Antennas Propagat.*, vol.5, no.4, pp.468-475, 2011.

[86] A.F. Molisch, "Ultrawideband Propagation Channel," in *Proc. IEEE*, 2009, vol.97, no.2, pp.353-371.

[87] A. Dezfooliyan and A.M. Weiner, "Phase Compensation Communication Technique against Time Reversal for Ultra-Wideband Channels," *IET Communications*, vol.7, no.12, pp.1287-1295, 2013.

[88] Y. Li and A.M. Weiner, "Photonic-assisted Error-free Wireless Communication with Multipath Pre-compensation covering 2-18 GHz," *IEEE T. Microw. Theory Tech.*, submitted, 2015.

[89] E. Hamidi and A.M. Weiner, "Phase-Only Matched Filtering of Ultrawideband Arbitrary Microwave Waveforms via Optical Pulse Shaping," *IEEE J. Lightw. Technol.*, vol.26, no.15, pp.2355-2363, 2008

[90] E. Hamidi and A.M. Weiner, "Post-compensation of Ultra-Wideband Antenna Dispersion using Microwave Photonic Phase Filters and Its Applications to UWB Systems," *IEEE T. Microw. Theory Tech.*, vol.57, no.4, pp.890 -898, 2009

[91] M.H. Song, V. Torres-Company , A.J. Metcalf, and A.M. Weiner "Multitap Microwave Photonic Filters with Programmable Phase Response via Optical Frequency Comb Shaping," *Opt. Lett.*, vol.37, no.5, pp.845-847, 2012.

[92] H.J. Kim, A. Rashidinejad, and A.M. Weiner, "Low-loss Ultrawideband Programmable RF Photonic Phase Filter for Spread Spectrum Pulse Compression," *IEEE T. Microw. Theory Tech.*, under review, 2015.

[93] J. Proakis and M. Salehi, *Digital Communications*: McGraw-Hill, 2007.

[94] A. Rashidinejad, D.E. Leaird, and A.M. Weiner, "Ultrabroadband Radio-Frequency Arbitrary Waveform Generation with High-Speed Phase and Amplitude Modulation Capability," *Opt. Express*, vol.23, no.9, pp.12265-12273, 2015.